\documentclass[twocolumn,showpacs,preprintnumbers,amsmath,amssymb,superscriptaddress,floatfix,altaffilletter]{revtex4-1}
\usepackage{graphicx}
\usepackage[pagewise,mathlines]{lineno}
\usepackage{bm}
\usepackage{indentfirst}
\usepackage{fancyhdr}
\usepackage{subfigure}
\usepackage{amssymb}
\usepackage{amsmath}
\usepackage{multirow}
\usepackage{comment}
\usepackage{rotating}
\usepackage{enumerate}
\usepackage{dcolumn}

\begin{document}

\title{\bf Measurement of $K^+$ production cross section by 8 GeV protons using high energy neutrino interactions in the SciBooNE detector}


\affiliation{Institut de Fisica d'Altes Energies, Universitat Autonoma de Barcelona, E-08193 Bellaterra (Barcelona), Spain}
\affiliation{Department of Physics, University of Colorado, Boulder, Colorado 80309, USA}
\affiliation{Department of Physics, Columbia University, New York, New York 10027, USA}
\affiliation{Fermi National Accelerator Laboratory; Batavia, Illinois 60510, USA}
\affiliation{High Energy Accelerator Research Organization (KEK), Tsukuba, Ibaraki 305-0801, Japan}
\affiliation{Department of Physics, Imperial College London, London SW7 2AZ, United Kingdom}
\affiliation{Department of Physics, Indiana University, Bloomington, Indiana 47405, USA}
\affiliation{Kamioka Observatory, Institute for Cosmic Ray Research, University of Tokyo, Gifu 506-1205, Japan}
\affiliation{Research Center for Cosmic Neutrinos, Institute for Cosmic Ray Research, University of Tokyo, Kashiwa, Chiba 277-8582, Japan}
\affiliation{Department of Physics, Kyoto University, Kyoto 606-8502, Japan}
\affiliation{Los Alamos National Laboratory, Los Alamos, New Mexico 87545, USA}
\affiliation{Department of Physics and Astronomy, Louisiana State University, Baton Rouge, Louisiana 70803, USA}
\affiliation{Department of Physics, Massachusetts Institute of Technology, Cambridge, Massachusetts 02139, USA}
\affiliation{Department of Chemistry and Physics, Purdue University Calumet, Hammond, Indiana 46323, USA}
\affiliation{Universit$\grave{a}$ di Roma Sapienza, Dipartimento di Fisica and INFN, I-00185 Rome, Italy}
\affiliation{Physics Department, Saint Mary's University of Minnesota, Winona, Minnesota 55987, USA}
\affiliation{Department of Physics, Tokyo Institute of Technology, Tokyo 152-8551, Japan}
\affiliation{Instituto de Fisica Corpuscular, Universidad de Valencia and CSIC, E-46071 Valencia, Spain}

\author{G.~Cheng}\affiliation{Department of Physics, Columbia University, New York, New York 10027, USA}
\author{C.~Mariani}\affiliation{Department of Physics, Columbia University, New York, New York 10027, USA}
\author{J.~L.~Alcaraz-Aunion}\affiliation{Institut de Fisica d'Altes Energies, Universitat Autonoma de Barcelona, E-08193 Bellaterra (Barcelona), Spain}
\author{S.~J.~Brice}\affiliation{Fermi National Accelerator Laboratory; Batavia, Illinois 60510, USA}
\author{L.~Bugel}\affiliation{Department of Physics, Massachusetts Institute of Technology, Cambridge, Massachusetts 02139, USA}
\author{J.~Catala-Perez}\affiliation{Instituto de Fisica Corpuscular, Universidad de Valencia and CSIC, E-46071 Valencia, Spain}
\author{J.~M.~Conrad}\affiliation{Department of Physics, Massachusetts Institute of Technology, Cambridge, Massachusetts 02139, USA}
\author{Z.~Djurcic}\affiliation{Department of Physics, Columbia University, New York, New York 10027, USA}
\author{U.~Dore}\affiliation{Universit$\grave{a}$ di Roma Sapienza, Dipartimento di Fisica and INFN, I-00185 Rome, Italy}
\author{D.~A.~Finley}\affiliation{Fermi National Accelerator Laboratory; Batavia, Illinois 60510, USA}
\author{A.~J.~Franke}\affiliation{Department of Physics, Columbia University, New York, New York 10027, USA}
\author{C.~Giganti}\altaffiliation[Present address: ]{DSM/Irfu/SPP, CEA Saclay, F-91191 Gif-sur-Yvette, France}\affiliation{Universit$\grave{a}$ di Roma Sapienza, Dipartimento di Fisica and INFN, I-00185 Rome, Italy}
\author{J.~J.~Gomez-Cadenas}\affiliation{Instituto de Fisica Corpuscular, Universidad de Valencia and CSIC, E-46071 Valencia, Spain}
\author{P.~Guzowski}\affiliation{Department of Physics, Imperial College London, London SW7 2AZ, United Kingdom}
\author{A.~Hanson}\affiliation{Department of Physics, Indiana University, Bloomington, Indiana 47405, USA}
\author{Y.~Hayato}\affiliation{Kamioka Observatory, Institute for Cosmic Ray Research, University of Tokyo, Gifu 506-1205, Japan}
\author{K.~Hiraide}\affiliation{Department of Physics, Kyoto University, Kyoto 606-8502, Japan}\affiliation{Kamioka Observatory, Institute for Cosmic Ray Research, University of Tokyo, Gifu 506-1205, Japan}
\author{G.~Jover-Manas}\affiliation{Institut de Fisica d'Altes Energies, Universitat Autonoma de Barcelona, E-08193 Bellaterra (Barcelona), Spain}
\author{G.~Karagiorgi}\affiliation{Department of Physics, Massachusetts Institute of Technology, Cambridge, Massachusetts 02139, USA}
\author{T.~Katori}\affiliation{Department of Physics, Indiana University, Bloomington, Indiana 47405, USA}
\affiliation{Department of Physics, Massachusetts Institute of Technology, Cambridge, Massachusetts 02139, USA}
\author{Y.~K.~Kobayashi}\affiliation{Department of Physics, Tokyo Institute of Technology, Tokyo 152-8551, Japan}
\author{T.~Kobilarcik}\affiliation{Fermi National Accelerator Laboratory; Batavia, Illinois 60510, USA}
\author{H.~Kubo}\affiliation{Department of Physics, Kyoto University, Kyoto 606-8502, Japan}
\author{Y.~Kurimoto}\affiliation{Department of Physics, Kyoto University, Kyoto 606-8502, Japan}\affiliation{High Energy Accelerator Research Organization (KEK), Tsukuba, Ibaraki 305-0801, Japan}
\author{W.~C.~Louis}\affiliation{Los Alamos National Laboratory, Los Alamos, New Mexico 87545, USA}
\author{P.~F.~Loverre}\affiliation{Universit$\grave{a}$ di Roma Sapienza, Dipartimento di Fisica and INFN, I-00185 Rome, Italy}
\author{L.~Ludovici}\affiliation{Universit$\grave{a}$ di Roma Sapienza, Dipartimento di Fisica and INFN, I-00185 Rome, Italy}
\author{K.~B.~M.~Mahn}\altaffiliation[Present address: ]{TRIUMF, Vancouver, British Columbia, V6T 2A3, Canada}\affiliation{Department of Physics, Columbia University, New York, New York 10027, USA}
\author{S.~Masuike}\affiliation{Department of Physics, Tokyo Institute of Technology, Tokyo 152-8551, Japan}
\author{K.~Matsuoka}\affiliation{Department of Physics, Kyoto University, Kyoto 606-8502, Japan}
\author{V.~T.~McGary}\affiliation{Department of Physics, Massachusetts Institute of Technology, Cambridge, Massachusetts 02139, USA}
\author{W.~Metcalf}\affiliation{Department of Physics and Astronomy, Louisiana State University, Baton Rouge, Louisiana 70803, USA}
\author{G.~B.~Mills}\affiliation{Los Alamos National Laboratory, Los Alamos, New Mexico 87545, USA}
\author{G.~Mitsuka}\altaffiliation[Present address: ]{Solar-Terrestrial Environment Laboratory, Nagoya University, Furo-cho, Chikusa-ku, Nagoya, Japan}\affiliation{Research Center for Cosmic Neutrinos, Institute for Cosmic Ray Research, University of Tokyo, Kashiwa, Chiba 277-8582, Japan}
\author{Y.~Miyachi}\altaffiliation[Present address: ] {Department of Physics, Yamagata University, Yamagata, 990-8560 Japan}\affiliation{Department of Physics, Tokyo Institute of Technology, Tokyo 152-8551, Japan}
\author{S.~Mizugashira}\affiliation{Department of Physics, Tokyo Institute of Technology, Tokyo 152-8551, Japan}
\author{C.~D.~Moore}\affiliation{Fermi National Accelerator Laboratory; Batavia, Illinois 60510, USA}
\author{Y.~Nakajima}\altaffiliation[Present address: ]{Lawrence Berkeley National Laboratory, Berkeley, CA 94720, USA}\affiliation{Department of Physics, Kyoto University, Kyoto 606-8502, Japan}
\author{T.~Nakaya}\affiliation{Department of Physics, Kyoto University, Kyoto 606-8502, Japan}
\author{R.~Napora}\affiliation{Department of Chemistry and Physics, Purdue University Calumet, Hammond, Indiana 46323, USA}
\author{P.~Nienaber}\affiliation{Physics Department, Saint Mary's University of Minnesota, Winona, Minnesota 55987, USA}
\author{D.~Orme}\affiliation{Department of Physics, Kyoto University, Kyoto 606-8502, Japan}
\author{M.~Otani}\affiliation{Department of Physics, Kyoto University, Kyoto 606-8502, Japan}
\author{A.~D.~Russell}\affiliation{Fermi National Accelerator Laboratory; Batavia, Illinois 60510, USA}
\author{F.~Sanchez}\affiliation{Institut de Fisica d'Altes Energies, Universitat Autonoma de Barcelona, E-08193 Bellaterra (Barcelona), Spain}
 \author{M.~H.~Shaevitz}\affiliation{Department of Physics, Columbia University, New York, New York 10027, USA}
\author{T.-A.~Shibata}\affiliation{Department of Physics, Tokyo Institute of Technology, Tokyo 152-8551, Japan}
\author{M.~Sorel}\affiliation{Instituto de Fisica Corpuscular, Universidad de Valencia and CSIC, E-46071 Valencia, Spain}
\author{R.~J.~Stefanski}\affiliation{Fermi National Accelerator Laboratory; Batavia, Illinois 60510, USA}
\author{H.~Takei}\altaffiliation[Present address: ]{Kitasato University, Tokyo, 108-8641 Japan}\affiliation{Department of Physics, Tokyo Institute of Technology, Tokyo 152-8551, Japan}
\author{H.-K.~Tanaka}\altaffiliation[Present address: ]{Brookhaven National Laboratory, Upton, New York 11973, USA}\affiliation{Department of Physics, Massachusetts Institute of Technology, Cambridge, Massachusetts 02139, USA}
\author{M.~Tanaka}\affiliation{High Energy Accelerator Research Organization (KEK), Tsukuba, Ibaraki 305-0801, Japan}
\author{R.~Tayloe}\affiliation{Department of Physics, Indiana University, Bloomington, Indiana 47405, USA}
\author{I.~J.~Taylor}\altaffiliation[Present address: ]{Department of Physics and Astronomy, State University of New York, Stony Brook, New York 11794-3800, USA }\affiliation{Department of Physics, Imperial College London, London SW7 2AZ, United Kingdom}
\author{R.~J.~Tesarek}\affiliation{Fermi National Accelerator Laboratory; Batavia, Illinois 60510, USA}
\author{Y.~Uchida}\affiliation{Department of Physics, Imperial College London, London SW7 2AZ, United Kingdom}
\author{R.~Van~de~Water}\affiliation{Los Alamos National Laboratory, Los Alamos, New Mexico 87545, USA}
\author{J.~J.~Walding}\altaffiliation[Present address: ]{Department of Physics, College of William \& Mary, Williamsburg, VA 23187, USA}\affiliation{Department of Physics, Imperial College London, London SW7 2AZ, United Kingdom}
\author{M.~O.~Wascko}\affiliation{Department of Physics, Imperial College London, London SW7 2AZ, United Kingdom}
\author{H.~B.~White}\affiliation{Fermi National Accelerator Laboratory; Batavia, Illinois 60510, USA}
\author{M.~Yokoyama}\altaffiliation[Present address: ]{Department of Physics, University of Tokyo, Tokyo 113-0033, Japan}\affiliation{Department of Physics, Kyoto University, Kyoto 606-8502, Japan}
\author{G.~P.~Zeller}\affiliation{Fermi National Accelerator Laboratory; Batavia, Illinois 60510, USA}
\author{E.~D.~Zimmerman}\affiliation{Department of Physics, University of Colorado, Boulder, Colorado 80309, USA}
\collaboration{SciBooNE Collaboration}\noaffiliation


\date{\today}


\begin{abstract}

The SciBooNE Collaboration reports $K^{+}$ production cross section and rate measurements using high energy daughter muon neutrino scattering data off the SciBar polystyrene (C$_8$H$_8$) target in the SciBooNE detector. The $K^{+}$ mesons are produced by 8 GeV protons striking a beryllium target in Fermilab Booster Neutrino Beam line (BNB). Using observed neutrino and antineutrino events in SciBooNE, we measure
$$\dfrac{d^2\sigma}{dpd\Omega}~= ~(5.34\pm0.76)~mb/(GeV/c \times sr) $$
for $p + Be\rightarrow K^+ + X$ at mean $K^{+}$ energy of 3.9~GeV and angle (with respect to the proton beam direction) of 3.7~degrees, corresponding to the selected $K^+$ sample. Compared to Monte Carlo predictions using previous higher energy $K^+$ production measurements, this measurement, which uses the NUANCE neutrino interaction generator, is consistent with a normalization factor of 0.85$\pm$0.12. This agreement is evidence that the extrapolation of the higher energy $K^{+}$ measurements to an 8 GeV beam energy using Feynman scaling is valid. This measurement reduces the error on the $K^+$ production cross section from 40\% to 14\%.

\end{abstract}

\maketitle


\section{Introduction}

Inclusive kaon production by low-energy protons (1 to 15 GeV) is of interest both theoretically and experimentally.  In this low-energy region, kaon production is dominated by exclusive processes.  For example, the lowest threshold $K^+$ production process is $p + p   \rightarrow K^+ +  \Lambda + p$, which for a fixed target setup has an incoming beam energy threshold of 2.52 GeV.  Since exclusive channel threshold effects are important, theoretical models such as Feynman scaling~\cite{Feyman::1969} may be better in describing low-energy production cross sections. Measurements of kaon production in this region are not extensive and do not cover wide kinematic regions. In addition, systematic data on the energy and target nuclei dependence is not available. Thus, new measurements of kaon production are needed in this region. Experimentally, kaon production is also relevant for neutrino experiments since important components of the incident neutrino flux come from kaon decays.
\par A primary motivation of this work is to verify the simulation of neutrinos from the Fermilab Booster Neutrino Beam (BNB) line with actual data. The BNB line provides neutrinos for the MiniBooNE~\cite{AguilarArevalo:2008yp} and SciBooNE~\cite{Hiraide:2006zq} experiments, as well as possible future experiments, including MicroBooNE~\cite{Soderberg:2009rz}. In this beam line, protons with 8 GeV kinetic energy are directed onto a 1.8-interaction length beryllium target. The average energy of $\pi^{+}$ ($K^{+}$) that decay to neutrinos in the MiniBooNE detector acceptance is 1.89 (2.66) GeV. Therefore 37.6\% (92.1\%) decay before the end of the 50 m long decay region. The relevant decay modes for MiniBooNE/SciBooNE are $\pi^{+}\ \rightarrow\mu^{+}\nu_{\mu}$, $K^{+}\ \rightarrow\mu^{+}\nu_{\mu}$, which produce 92.9\% of the neutrino beam, $\pi^{-}\ \rightarrow\mu^{-}\bar{\nu}_{\mu}$, which produces 6.5\% of the neutrino beam, and $K^{+}\rightarrow\pi^{0} e^{+}\nu_{e}$, $\mu^{+}\rightarrow e^{+}\bar{\nu}_{\mu}\nu_{e}$, $K_{L}^{0}\rightarrow\pi^{-}e^{+}\nu_{e}$, and $K_{L}^{0}\rightarrow\pi^{+}e^{-}\bar{\nu}_{e}$, which produce 0.6\% of the neutrino beam.

While the neutrino flux is predominantly due to $\pi^{+}$ decay, $K^{+}$ decay is the dominant source above 2 GeV. The neutrinos from kaons provide a unique source of high energy events for experiments on the BNB line studying neutrino cross sections, and can represent a source of background for experiments exploring neutrino oscillations and beyond-the-standard-model effects.  Therefore, it is important for the BNB line experiments to understand the rate of $K^{+}$ production.

An accurate understanding of $K^{+}$ production will reduce systematics associated with the measured $\nu_e$ background in MiniBooNE, a major contributor to the uncertainty in the previously published $\nu_e$ oscillation appearance result~\cite{AguilarArevalo:2009xn}.
The measurement of $K^{+}$ production in this energy region combined with $K^{+}$ production at higher energies is a good test of production models such as the Feynman-Scaling~\cite{shaevitz:2010} and Modified Sanford-Wang~\cite{Sanford:1967} parameterization used to describe secondary meson production at low primary proton beam energy in the BNB.

This work describes a measurement of $K^{+}$ production by measuring the rate of high-energy $\nu_\mu$ events from kaon decay. The data sample used for the measurement comes from the interaction of $\nu_{\mu}$ and $\bar{\nu}_\mu$ which undergo charged current (CC) neutrino interactions in the fiducial volume of the SciBooNE detector, generating high energy $\mu^-$ and $\mu^+$ (along with a host of other particles) that penetrate the entire SciBooNE detector, providing essentially a minimum muon momentum requirement of 1.0 GeV/c. The neutrinos from $K^{+}$ decay can be isolated using the angular distribution of the outgoing muons and the multiplicity of charged particles produced in the interaction. The number of $K^{+}$ decay neutrinos is then compared to prediction to make a determination of a normalization factor for production in the BNB and correspondingly a $K^{+}$ production cross section. The paucity of high-energy events in the SciBooNE experiment prevents a measurement of the kinematic distribution of $K^{+}$ production but does allow a normalization determination with improved uncertainty.


\section{SciBooNE Experiment}\label{sec:experiment}

\subsection{Neutrino Beam}\label{sec:beam}

The SciBooNE experiment detected neutrinos produced by the Fermilab BNB. The same BNB beam is also serving the MiniBooNE experiment. The BNB uses protons accelerated to 8 GeV kinetic energy by the Fermilab Booster synchrotron. Beam properties are monitored on a spill-by-spill basis, and at various locations along the BNB line. Transverse and directional alignment of the beam, beam width and angular divergence, beam intensity and losses along the BNB, are measured and used in the data quality selection. Protons strike a 71.1 cm long beryllium target, producing a secondary beam of hadrons, mainly pions with a small fraction of kaons. A cylindrical horn electromagnet made of aluminum surrounds the beryllium target to sign-select and focus the secondary beam. For neutrino running mode, the horn polarity was set to focus particles with positive electric charge and for antineutrino running mode, the horn polarity was set to focus particles with negative electric charge. The neutrino beam is mostly produced in the 50 m long decay region. The analysis described in this paper will use data from both neutrino and antineutrino running modes.

The Monte Carlo (MC) simulation of the neutrino beam was modeled by the MiniBooNE collaboration. The MiniBooNE collaboration
uses a GEANT4-based Monte Carlo simulation that can be roughly divided into five consecutive simulation steps. The first simulation
step is the definition of the beam-line geometry including the shape, location, and composition of components. The second simulation
step is the generation of primary protons according to the measured beam optics properties. The third simulation step is the simulation of
particles produced by the initial p-Be interaction, including the elastic and quasi-elastic scattering of the protons in the target.
Custom tables for the production of proton, neutron, $\pi^{\pm}$, $K^{\pm}$, $K^{0}$ are used based on the phenomenology of particle production
and data of the production of these particles at higher energies.
The fourth simulation step is the propagation of the produced particles using the
GEANT4 framework taking into account energy loss, electromagnetic and hadronic processes, and trajectory deflection by the magnetic field created
by the horn. The fifth simulation step is the decay of propagated particles into neutrinos using the current branching fraction measurements~\cite{AguilarArevalo:2008yp}.

Particle production is simulated using the methods described in Ref.~\cite{AguilarArevalo:2008yp}.  The production of $K^{+}$ is simulated using a Feynman scaling formalism based on $K^{+}$ p-Be production data at different primary proton energies~\cite{AguilarArevalo:2008yp,shaevitz:2010}. The predicted double differential cross section using the Feynman parametrization reported in~\cite{AguilarArevalo:2008yp,shaevitz:2010} is

\begin{eqnarray}
\dfrac{d^2\sigma}{dpd\Omega}~= ~(6.3\pm2.5)~mb/(GeV/c \times sr),
\label{eq:central_value_doublediff}
\end{eqnarray}

at the mean $K^{+}$ energy of 3.9~GeV and mean angle of 3.7~degrees, which are the mean energy and angle for kaons which produce neutrinos in SciBooNE.

For $\pi^+$ and $\pi^-$ production, the Sanford-Wang (SW)~\cite{Sanford:1967} parametrization to the HARP p-Be data~\cite{:2007gt} at 8.89 GeV/c is used to determine the central value with associated errors determined from spline fits. SW~\cite{Sanford:1967} production is also used for $K^0$ production and errors. The long life time of the $K^0_L$ combined with the fact that they are not focused by the magnetic horn leads to the expectation that the contribution of decay neutrinos for this source is small relative to the $K^{+}$. For $K^{-}$ production, the scarcity of production measurements in the relevant kinematic regions motivated the use of the MARS hadronic interaction package~\cite{Mokhov:1998kc} to determine the absolute double differential cross-sections.

The neutrinos produced from the simulation are extrapolated along straight lines toward the SciBooNE detector. All neutrinos whose ray traces cross any part of the detector volume and surrounding rock are considered in the SciBooNE neutrino flux prediction and the kinematics of the neutrino and their parents are stored.

In the neutrino mode running (positive horn polarity), a total neutrino flux of $2.2\times 10^{-8}$cm$^{-2}$/POT is expected at the SciBooNE detector location, with a mean neutrino energy of 0.7 GeV. The flux is dominated by muon neutrinos (92.92\% of total), with small contributions from muon antineutrinos (6.48\%),  electron neutrinos (0.54\%) and electron antineutrinos (0.05\%).

In the antineutrino mode running (negative horn polarity), a total neutrino flux of $1.3\times 10^{-8}$cm$^{-2}$/POT is expected at the SciBooNE detector location, with a mean neutrino energy of 0.6 GeV. The flux is dominated by muon antineutrinos (83.85\% of total), with contributions from muon neutrinos (15.58\%), electron neutrinos (0.15\%) and electron antineutrinos (0.42\%).

\begin{figure}[htbp!]
\begin{center}
\subfigure{\includegraphics[width = \columnwidth]{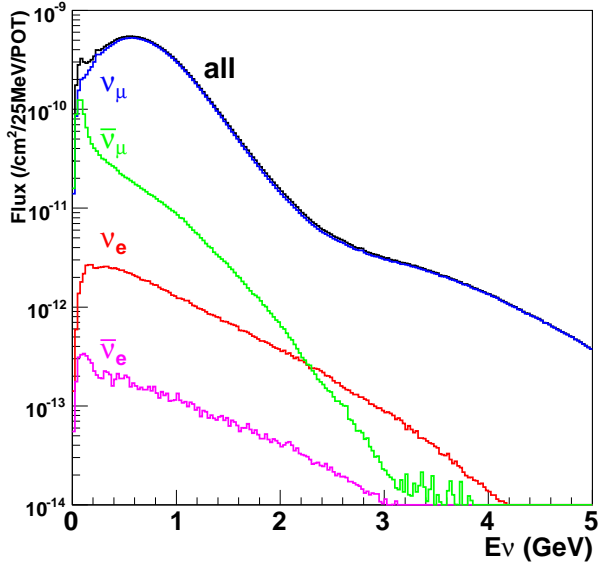}}
\subfigure{\includegraphics[width = \columnwidth]{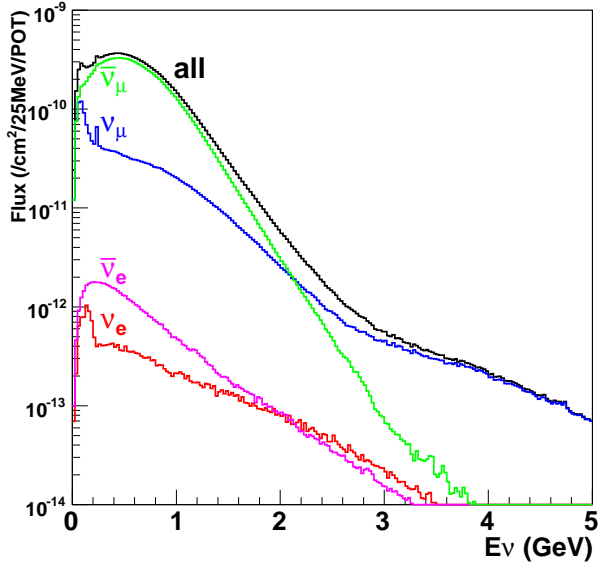}}
\caption{Neutrino flux predictions at the SciBooNE detector as a function of neutrino energy $E_{\nu}$, normalized per unit area, proton on target (POT), and neutrino energy bin width, in neutrino (top) and antineutrino (bottom) modes. The spectra is averaged within a circle with radius 2.12m from beam center (coincides with center of detector cross-sectional area), which covers the entire 3m$\times$3m cross-sectional area of the SciBooNE detector. The total flux and contributions from individual neutrino flavors are shown.}
\label{fig:nuflux}
\end{center}
\end{figure}

The neutrino flux predictions at the SciBooNE detector location as a function of neutrino energy for both neutrino and antineutrino mode running are shown in Fig.~\ref{fig:nuflux}. For the low-energy BNB, the neutrino spectrum at SciBooNE and MiniBooNE are very similar, except for the flux normalization difference \cite{Nakajima:2011zza}, in particular, according to simulations, the mean energy of the $\nu_\mu$ flux is expected to be 0.76 (0.79) GeV at the SciBooNE (MiniBooNE) detector location, reflecting the very similar fraction of $\nu_\mu$s from $\pi^+$ and $K^+$ decay at the two locations.

\begin{figure}[bhtp!]
\center
  \includegraphics[width=\columnwidth]{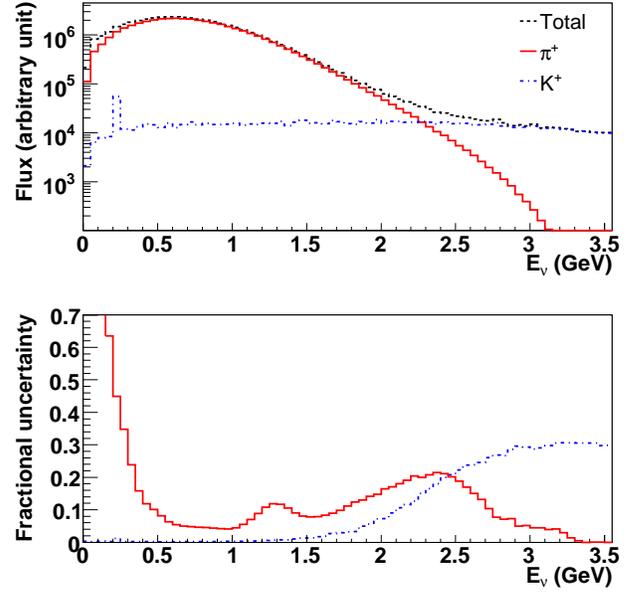}
  \caption{
    (Top) $\nu_\mu$ flux prediction at the SciBooNE detector as a function of neutrino energy $E_{\nu}$ in neutrino running mode.
    The total flux and contributions from $\pi^+$ and $K^+$ decays are shown.
    (Bottom) Fractional uncertainties of the $\nu_\mu$ flux prediction due to $\pi^+$ and $K^+$ production from the p-Be interaction in neutrino running mode.
    The figures are from Ref.~\cite{Nakajima:2010fp}.
  }
  \label{fig:numu_flux-parents}
\end{figure}

\begin{figure}[bhtp!]
\center
  \includegraphics[width=\columnwidth]{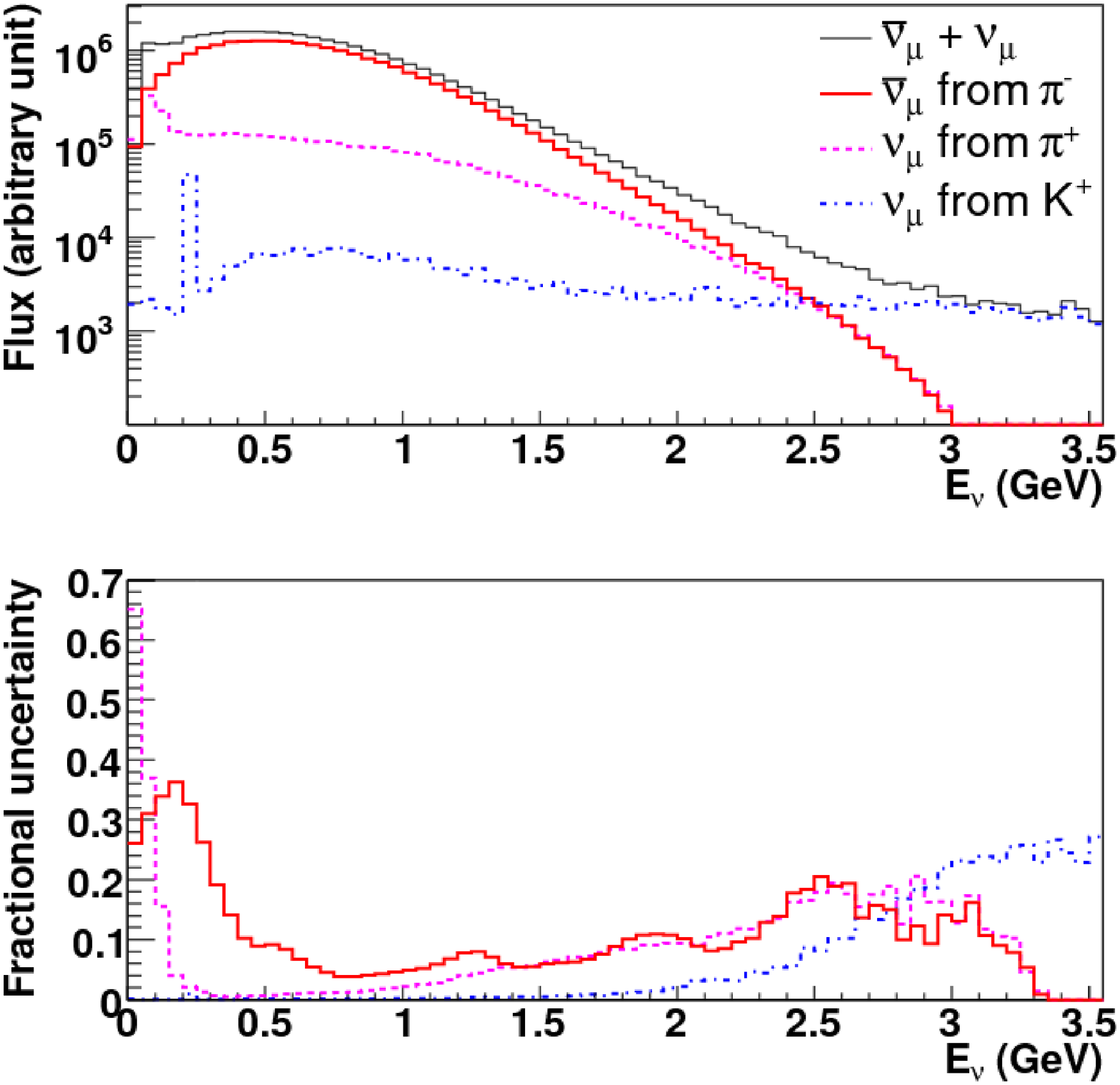}
  \caption{
    (Top) $\nu_\mu$ and $\bar{\nu}_\mu$ flux predictions at the SciBooNE detector as a function of neutrino energy $E_{\nu}$ in antineutrino running mode.
    The total flux and contributions from $\bar{\nu}_\mu$ from $\pi^-$ decay and $\nu_\mu$ from $\pi^+$ and $K^+$ decays are shown.
    (Bottom) Fractional uncertainties of the $\nu_{\mu}$ and $\bar{\nu_{\mu}}$ flux predictions due to $\pi^-$, $\pi^+$ and $K^+$ production from
    p-Be interactions in antineutrino running mode with respect to the total $\nu_{\mu} + \bar{\nu_{\mu}}$ flux.
  }
  \label{fig:antinumu_flux-parents}
\end{figure}

The systematic uncertainties in the neutrino flux prediction come from several sources in the simulation steps above: proton delivery/optics, secondary particle productions, hadronic interactions in the target or horn and horn magnetic field. Fig.~\ref{fig:numu_flux-parents} shows the $\nu_\mu$ flux in neutrino mode from $\pi^+$ and $K^+$ decays, and their fractional uncertainties. Fig.~\ref{fig:antinumu_flux-parents} shows the $\nu_\mu$ and $\bar{\nu}_\mu$ flux in antineutrino mode from $\pi^-$, $\pi^+$, and $K^+$ decays, and their fractional uncertainties. The uncertainty in $\pi^+$ ($\pi^-$) production is determined from uncertainties associated with spline fits to the HARP $\pi^+$ ($\pi^-$) double differential cross section data~\cite{AguilarArevalo:2008yp} and differences between the SW and spline fit central values. The HARP data used were those from a thin (5\% interaction length) beryllium target run~\cite{:2007gt}. While the HARP data provide a valuable constraint on the BNB flux prediction, additional uncertainties resulting from thick target effects (secondary re-scattering of protons and pions) are included through the BNB flux simulation. The resulting $\pi^+$ production uncertainty is $\approx$ 5\% at the peak of the flux distribution and increases significantly at high and low neutrino energies. The resulting $\pi^-$ production uncertainty is $\approx$ 10\% at the peak of the flux distribution and also increases at high and low neutrino energies.

The flux from $K^+$ decay is dominant for $E_\nu > 2.0$ GeV. Since no published data exist for $K^+$ production at the BNB primary
proton beam energy, we employ the Feynman scaling hypothesis to relate $K^+$ production measurements at different proton beam energies to the expected production at the BNB proton beam energy~\cite{AguilarArevalo:2008yp}. The errors of the Feynman scaling parameters obtained from these measurements are then included as systematic uncertainties.

Other major contributions to the flux error include uncertainties in the hadron interactions at the target and simulation of the the horn magnetic field, which both contribute to shape and normalization uncertainties. An overall normalization uncertainty is included on the number of protons on target (POT). All flux errors are modeled through variations in the simulation and result in a total error of $\approx$ 7\% at the peak of the flux. Quantitative constraints of each uncertainty have been determined from previous MiniBooNE studies~\cite{AguilarArevalo:2008yp}.

\subsection{SciBooNE Detector}

\begin{figure}[bhtp!]
\center
  \includegraphics[width=\columnwidth]{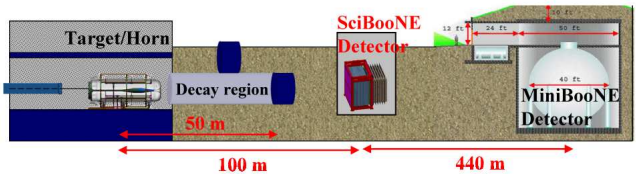}
  \caption{Schematic overview of the Booster Neutrino Beamline
    and the location of the SciBooNE and the MiniBooNE detectors.}
  \label{fig:sciboone_setup}
\end{figure}

The SciBooNE detector was located 100~m downstream from the beryllium target on the axis of the beam, as shown in Fig.~\ref{fig:sciboone_setup}.
The detector was comprised of three sub-detectors: a fully active and finely segmented scintillator
tracker (SciBar), an electromagnetic calorimeter (EC), and a muon range detector (MRD). SciBar served as the primary neutrino target
for this analysis.

SciBooNE uses a right-handed Cartesian coordinate system in which the $z$ axis is the beam direction and the $y$ axis is the vertical upward direction. The origin is located on the most upstream surface of SciBar in the $z$ dimension, and at the center of the SciBar scintillator plane in the $x$ and $y$ dimensions. Since each sub-detector is read out both vertically and horizontally, two views are defined: top ($x$~vs.~$z$ projection) and side ($y$~vs.~$z$ projection).

The SciBar detector~\cite{Nitta:2004nt} was positioned upstream of the other sub-detectors. It consists of 14,336 extruded plastic scintillator strips. Each strip has a dimension of 2.5 $\times$ 300 $\times$ 1.3~cm$^3$. The scintillators are arranged vertically and horizontally to construct a 3 $\times$ 3 $\times$ 1.7~m$^3$ volume with a total mass of 15 tons. The dominant component of the SciBar detector is polystyrene ($\rm C_8H_8$). The uncertainty of the total detector mass is estimated to be 1\%, including the effect of epoxy resin used to glue the strips.

Each strip was read out by a 64-channel multi-anode photo-multiplier (MA-PMT) via a wavelength shifting (WLS) fiber.  Charge information was recorded for each channel, while timing information was recorded in groups of 32 channels by taking the logical OR with multi-hit TDC modules~\cite{Yoshida:2004mh}. The timing resolution for minimum-ionizing particles was evaluated with cosmic ray data to be 1.6~ns. The average light yield for minimum-ionizing particles is approximately 20 photo-electrons per 1.3~cm path length, and the typical pedestal width is below 0.3 photo-electron. The hit finding efficiency evaluated with cosmic ray data is more than 99.8\%. The minimum length of a reconstructable track is approximately 8~cm (three layers hit in each view). The track finding efficiency for single tracks of 10~cm or longer is more than 99\%. Large samples of events were visually scanned to ensure that the track finding algorithm is efficient, has high purity, and is unbiased. The scanning showed that the purity of the track-finding algorithm using the information available from the detector was efficient for separating events into SciBar 1, 2, 3-Track samples.

The EC is located just downstream of SciBar, and is designed to measure the electron neutrino contamination in the beam and tag photons from $\pi^0$ decay. The EC is a ``spaghetti" type calorimeter comprised of 1~mm diameter scintillating fibers embedded in lead foil~\cite{Mariani:2009zza}.  The calorimeter is made of 64 modules of dimensions 262 $\times$ 8 $\times$ 4~cm$^3$. The fibers are bundled in two independent groups of 4 $\times$ 4~cm$^2$ transverse cross section, read at both ends by Hamamatsu PMTs.

The MRD was installed downstream of the EC and is designed to measure the momentum of muons produced by CC neutrino interactions. It comprised of 12 iron plates with thickness 5~cm sandwiched between planes of 6~mm thick scintillation counters; there were 13 alternating horizontal and vertical planes read out via 362 individual 2~inch PMTs. Each iron plate measured 274 $\times$ 305~cm$^2$. The MRD measured the momentum of muons up to 1.2~GeV/$c$ using the observed muon range. Charge and timing information from each PMT were recorded. The average hit finding efficiency is 99\%.


\subsection{Detector Response Simulation}

The GEANT4 framework is used for the detector simulation. The Bertini cascade model within GEANT4~\cite{Heikkinen:2003sc} is used to simulate
the interactions of hadronic particles with detector materials. The detector simulation includes a detailed geometric model of the detector, including the detector frame, experimental hall, and soil. The energy loss of a particle in each single SciBar strip and each
individual EC sensitive fiber is simulated. The energy deposition is converted in the detector response taking into account the Birk's saturation of the scintillator, the light attenuation along the fibers, the Poisson fluctuation of the number of photoelectrons, the PMT resolution, and electronic noise. The crosstalk in nearby SciBar channels is also simulated.

In SciBar the timing of each hit is simulated from the true time of the corresponding energy deposition, corrected by the travel time of the
light in the WLS fiber and smeared by the timing resolution.

For the detector simulation of the MRD, true energy deposition in each scintillator is converted to ADC counts using the conversion factor measured with cosmic muons. The attenuation of light in the scintillator as well as electronics noise are simulated. The time of energy deposition is digitized and converted into TDC counts.

The input parameters of the detector simulation are derived from laboratory measurements and calibration data. The features of the simulation have been systematically compared and tuned with cosmic ray and neutrino data.

A more detailed description of the detector simulation is given in~\cite{Hiraide:2008eu}.


\subsection{Neutrino Interaction Simulation}\label{sec:neutrinointeractionsimulation}

In the SciBooNE experiment, neutrino interactions with carbon and hydrogen in the SciBar detector
are simulated by the NEUT~\cite{Hayato:2002sd,Mitsuka:2008zz} and NUANCE~\cite{Casper:2002sd} program libraries.
NEUT is used in Kamiokande, Super-Kamiokande, K2K, and T2K experiments, while NUANCE is used in MiniBooNE.
Several Monte Carlo samples with different NEUT and NUANCE implementations are produced, and compared to the SciBooNE neutrino data.
For the analysis presented in this paper we use NUANCE as the neutrino interaction simulation code because this Monte Carlo matches what was used in the MiniBooNE oscillation analysis~\cite{AguilarArevalo:2007it}. The same analysis will also be repeated using NEUT and presented in Appendix~\ref{sec:neutnuance}. The total number of protons on target (POT) collected in neutrino mode is $0.99 \times 10^{20}$ while the POT for the antineutrino mode is $1.51 \times 10^{20}$. The expected number of events in the SciBooNE detector for each neutrino (antineutrino) interaction is listed in Tab.~\ref{tab:nuance}.

\begin{table}[htbn!]
\begin{center}
\caption{The expected number of events in each neutrino interaction estimated by NUANCE at the SciBooNE detector location with the neutrino beam exposure of $0.99 \times 10^{20}$ protons on target for neutrino mode and of $1.51 \times 10^{20}$ for antineutrino mode. The 9.8~ton fiducial volume of the SciBar detector is assumed. Charged Current and Neutral Current interactions are abbreviated as CC and NC, respectively.}
\vspace{0.5cm}
 \begin{tabular}{lcc}\hline\hline
                                       &  \multicolumn{2}{c}{NUANCE} \\
                                       &  \# Events ~~~~& \# Events \\
   Interaction                   & neutrino mode ~~~~& antineutrino mode\\
   \hline
    CC QE                        &  45,163 ~~~~  &   15,361         \\
    CC single-$\pi$              &  24,437 ~~~~ &   6,413          \\
    CC coherent $\pi$            &   1,706 ~~~~ &   1,326          \\
    CC DIS+Other                 &   3,049 ~~~~ &   1,518          \\
    NC                           &  29,118 ~~~~ &   11,686          \\\hline\hline
 \end{tabular}
\label{tab:nuance}
\end{center}
\end{table}

For neutrino mode beam exposure, the total number of CC interactions predicted by NUANCE integrated over the SciBooNE flux in the 9.8 ton SciBar fiducial volume is $7.44 \times 10^4$ for  $0.99 \times 10^{20}$ POT. For antineutrino mode beam exposure, the total number of CC interactions predicted by NUANCE integrated over the SciBooNE flux in the 9.8 ton SciBar fiducial volume is $2.46 \times 10^4$ for $1.51 \times 10^{20}$ POT.

The targets handled in NUANCE are proton, neutron, and carbon nuclei. The types of neutrino interactions simulated in both Neutral Current (NC) and CC are: elastic and quasi-elastic
scattering ($\nu N \rightarrow \ell N'$), single meson production ($\nu N \rightarrow \ell N'm$), single gamma production ($\nu N
\rightarrow \ell N' \gamma$), coherent $\pi$ production, and deep inelastic scattering ($\nu N \rightarrow \ell N'hadrons$), where $N$ and $N'$ are the nucleons (proton or
neutron), $\ell$ is the lepton (electron, muon or neutrino), and $m$ is the meson. NUANCE also simulates resonantly-produced multi-pion and kaon production ($K^+ \Lambda$, $K^+ \Sigma$). In nuclei, interactions of the mesons and hadrons with the nuclear medium are simulated following the neutrino interactions.

In addition to neutrino interactions inside SciBar, we also simulate interactions in the EC and MRD and the surrounding materials (the walls of the detector hall and soil).


\subsubsection{Quasi-elastic Scattering}\label{ssec:qesiml}

The dominant interaction in SciBooNE is CC quasi-elastic scattering, which is implemented using the Smith and Moniz model~\cite{Smith:1972xh}.
The nucleons are treated as quasi-free particles with Fermi motion and Pauli blocking taken into account. The Fermi surface momentum ($p_F$) for carbon is set to 220~MeV/c and the nuclear potential ($E_B$) is set to 34~MeV/c, as extracted from electron scattering data~\cite{Moniz:1971mt} taking into account neutrino vs. electron scattering differences~\cite{AguilarArevalo:2008fb}. For the vector form factor, NUANCE uses the BBA-2003 form factor~\cite{Budd:2003wb} with a dipole form for the axial form factor with an adjustable axial mass, $M_A^{QE}$ for Carbon ($M_A^{QE}$-C) and for Hydrogen ($M_A^{QE}$-H). The $M_A^{QE}$-H is used only in the antineutrino mode running. In NUANCE, an empirical Pauli-blocking parameter, $\kappa$ is introduced~\cite{AguilarArevalo:2010zc} to better describe the MiniBooNE quasi-elastic data at low momentum transfer. When $\kappa > 1$, the Pauli-blocking of final state nucleons is increased and hence the cross section at low momentum transfer is suppressed. The values of ($M_A^{QE}$-C = 1.234, $M_A^{QE}$-H = 1.0) $\rm GeV/c^2$ and $\kappa = 1.022$ are used~\cite{AguilarArevalo:2008fb}. The same Fermi momentum distribution and nuclear potential are used in all other neutrino-nucleus interactions except for coherent $\pi$ production.

\subsubsection{Meson Production via Baryon Resonances}\label{ssec:spisim}

The second most frequent interaction in SciBooNE is the resonant production of single pion, kaon, and eta mesons which NUANCE described with the model of Rein and Sehgal~\cite{Rein:1980wg}. The Rein and Sehgal model assumes an intermediate baryon resonance, $N^*$, in the reaction of $\nu N \rightarrow \ell N^*, N^* \rightarrow
N'm$. All intermediate baryon resonances with mass less than 1.7~GeV/$c^2$ are included. Interactions with invariant masses greater than 2~GeV/$c^2$ are simulated as deep inelastic scattering. $\Delta$ re-interactions ($\Delta N \to NN$) which do not lead to a mesonic final state are also simulated. This re-interaction probability is assumed to be 10\% (20\%) for $\Delta^{++/-}$ ( $\Delta^{+/0}$) resonances. To determine the angular distribution of final state pions, the method of~\cite{Rein:1987cb} is used for the $P_{33}(1232)$ resonance. For other resonances, the directional distribution of the generated pion is chosen to be isotropic in the resonance rest frame. The axial-vector form factors are formalized to be dipole with $M_A^{1\pi}=1.10$~$\rm GeV/c^2$.

\subsubsection{Coherent Pion Production}\label{ssec:cohpisim}

Coherent pion production is a neutrino interaction with a nucleus which remains intact, releasing one pion with the same charge as the incoming weak current.
Because of the small momentum transfer to the target nucleus, the outgoing pion tends to be emitted in the forward direction, closely following the incoming neutrino direction. The formalism developed by Rein and Sehgal~\cite{Rein:1982pf,Rein:2006di} is used to simulate such interactions. The axial vector mass, $M_A^{coherent}$, is set to $1.03 \pm 0.28$~GeV/$c^2$. The total and inelastic
pion-nucleon cross sections from the original Rein-Sehgal publications~\cite{Rein:1982pf,Rein:2006di} are obtained from fits to PDG data and implemented as a function of pion energy. The Rein and Sehgal model predicts the CC coherent $\pi^+$ production rate to be approximately 1\% of the total neutrino CC rate in SciBooNE.

\subsubsection{Deep Inelastic Scattering}\label{ssec:delsim}

The deep inelastic scattering (DIS) cross section is calculated using the GRV98 parton distribution functions~\cite{Gluck:1998xa}.
Additionally, we have included the corrections in the small $Q^2$ region developed by Bodek and Yang~\cite{Bodek:2003wd}.
The DIS contribution slowly increases for W values starting at 1.7 GeV and becomes the only source of neutrino interactions above W$>$2 GeV. This is done to create a smooth transition between the resonance and DIS models and ensure continuity in distributions of kinematics and hadron multiplicity in the region of overlap.
Tab.~\ref{tab:xsec_params} summarizes the parameter choices used in NUANCE.

\begin{table}[htbp!]
  \begin{center}
  \caption{NUANCE parameters used for neutrino interaction simulation.}
  \vspace{0.5cm}
  \begin{tabular}{cc}\hline\hline
    $p_{F}$   & 220 MeV\\
    $E_{B}$   &  34 MeV \\
    $M_A^{QE}$-C     & 1.234 GeV\\
    $M_A^{QE}$-H     & 1.0 GeV\\
    $\kappa$  &  1.022    \\
    $M_A^{1\pi}$ &1.10 GeV\\
    $M_A^{coherent}$ &1.03 GeV\\
    $M_A^{N\pi}$ &1.3 GeV\\\hline\hline
  \end{tabular}
  \label{tab:xsec_params}
  \end{center}
\end{table}

\subsubsection{Intra-nuclear Interactions}\label{ssec:fsisim}

Following production, the intra-nuclear interactions of mesons and nucleons are simulated using a cascade model in which the particles are traced until they escape the target nucleus.

Although we only use kinematic information from the final state muon in this analysis, the simulation of internuclear interaction is important since the pions/protons emitted from the nucleus can affect the selection criteria and bias the kinematics of the final state muon.

Inelastic scattering, charge exchange and absorption of pions in nuclei are simulated. For inelastic scattering and charge exchange interactions, the direction and momentum of pions are affected. In the scattering amplitude, Pauli blocking is also taken into account. A more detailed description of the intra-nuclear interaction simulations in NUANCE can be found in~\cite{Casper:2002sd}.


\section{Neutrino Mode Analysis}\label{sec:muon_neutrino_analysis}

In this analysis, we use a sample of high energy $\nu_\mu$ events identified by muons penetrating the SciBar, EC, and MRD detectors. According to our Monte Carlo simulation, 43\% of these high energy neutrinos come from $K^{+}$ decays in the beam.

The entire neutrino-mode data set for the SciBooNE experiment is used for this analysis. The neutrino run occurred from October 2007 to April 2008. A total of $0.99 \times 10^{20}$ POT, after all data quality cuts, was collected with an efficiency of 95\%.

The hit threshold in SciBar is set at 2.5 photoelectrons (corresponding to roughly 0.25 MeV) for a scintillator strip. The hits in each view are then associated into two-dimensional (2D) tracks in each view using a cellular automaton algorithm developed in K2K~\cite{Glazov:1993ur}. First, a correction for cross talk is applied to both data and MC prediction before 2D track reconstruction. Then, the remaining hits are categorized into ``clusters", where a cluster is one or more hits in adjacent scintillator strips. Segments are formed, which connect individual clusters no more than one scintillator strip apart. Any segments which share a cluster are then connected, provided the $\chi^2$ of a least squares linear fit remains acceptable. Three dimensional tracks are formed from 2D tracks in both views by requiring the timing between the 2D tracks to be within 50 ns, and the start and end point in the z direction for the 2D tracks to be within 6.6 cm.

A reconstructed muon track (referred to as a SciBar-MRD matched track) consists of a SciBar reconstructed track matching a reconstructed track in the MRD. The fiducial volume (FV) is applied requiring the upstream edge of the track to be within $|x| \leq 130$ cm, $|y| \leq 130$ cm, and $5.24 \leq |z| \leq 149.34$ cm. This corresponds to a 9.7 m$^3$ FV and 9.8 tons of fiducial mass. This selection criterion is not the same as is used in previous SciBooNE publications of 10.6 tons \cite{Hiraide:2009zz,Kurimoto:2010zz,Nakajima:2010fp}; the selection has been optimized for the particular analysis described here. The difference is due to a more conservative cut in the z direction to better reduce background events and more accurately reconstruct kinematic variables associated with the SciBar tracks.

As only MRD events are selected, a muon-hypothesis track selection is unnecessary, and the FV selection has been modified to boost the number of events in the sample without adding backgrounds. The track is also required to be in time with the neutrino beam, i.e. within $0 \leq t \leq 2,000$ ns. The extrapolation of the SciBar track is required to be matched with a track in MRD within 30 cm and within a 100 ns time difference. The matching requires a track that penetrates at least four consecutive MRD layers. If no MRD track is found, MRD hits can be used for the matching. The MRD hits, which in this case are required to be within a cone with an aperture of $\pm0.5$ rad, should have a time within 100 ns with respect to the SciBar track and should be within 10 cm of the extrapolated SciBar track on the MRD first layer.

The SciBar-MRD matched track can be further classified based on the end point of the MRD matched track: if the SciBar-MRD matched track stops in the MRD, the track is defined as SciBar-MRD stopped. If the track goes through all of the MRD planes, the track is classified as SciBar-MRD penetrated. If the SciBar-MRD matched track escapes from the side of the MRD before the most downstream plane, the event is classified as SciBar-MRD side-escaped.

In this analysis we will consider only events with one SciBar-MRD penetrated track. This defines a sample enriched in high energy muon neutrino events. The selected muons will have a minimum momentum of 1.0 GeV/c and muon angle relative to beam axis of less than 45 degrees.

The data sets for SciBooNE are subject to contamination by cosmic backgrounds (due to lack of detector shielding or veto). The vast majority of the cosmic backgrounds come from cosmic muons, which can mimic a muon signal from a $\nu_\mu$ interaction in SciBar. The selection cuts remove most of the cosmic background muons due to differences between cosmic background muons and muons created by neutrino beam interactions. The FV and timing cuts in SciBar remove the cosmic backgrounds coming from outside the detector and beam window. Most of the cosmic backgrounds enter the detector from above, producing vertical tracks (neutrino beam event tracks are generally horizontal). The vertical cosmic tracks do not pass through both SciBar and MRD and no SciBar-MRD matched track can be reconstructed, failing the cut requiring a SciBar-MRD matched track. For events with a SciBar-MRD matched track, the cosmic background contamination in the beam timing window is 0.5\%, estimated using a beam-off timing window ($5,000 \leq t \leq 15,000$ ns). To further correct for the cosmic backgrounds, identical selection cuts are made in data in the timing region $5,000 \leq t \leq 15,000$ ns, far outside the beam timing window and 5 times as long for better statistics. The data numbers and distributions selected in this cosmic background region are then divided by 5 and subtracted from the corresponding data events and distributions selected in the beam timing window. The resulting cosmic background subtracted data numbers and distributions assume no further cosmic background contamination. There is no cosmic background modeling in the MC. The number of events from cosmic backgrounds, neutrino interactions in the EC and MRD, and events with neutrino interactions in the material surrounding the SciBar detector are negligible. For all three samples combined, the number of background events is estimated to be 26 events: 15 from cosmic, 9 from backscattering neutrino events in EC/MRD, and 2 from interaction outside the SciBooNE detectors.

The high energy muons penetrate through the entire SciBooNE detector so the reconstruction of the total muon energy could not be done. The reconstructed muon angle relative to beam axis will be used as the primary kinematic variable for the analysis. Neutrinos produced from $K^+$ decay have a higher energy on average than neutrinos from $\pi^+$ decay. Therefore, the angular distribution of the resulting muon from the neutrino interaction of a neutrino from $K^+$ will be more forward peaked than those from neutrinos from $\pi^+$.

\begin{figure}[htbn!]
\begin{center}
\includegraphics[width=\columnwidth]{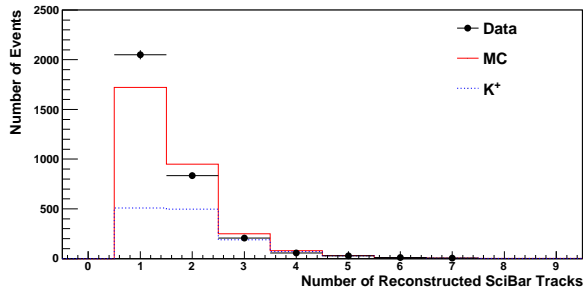}
\caption{Number of SciBar reconstructed tracks for the selected event sample in neutrino mode for data and NUANCE MC. MC histogram includes all events and $K^+$ shows the component of $\nu_\mu$ from $K^+$.}
\label{fig:selectedsample_ntracks}
\end{center}
\end{figure}

\begin{figure}[htbn!]
\begin{center}
\subfigure[~1-Track Sample]{\includegraphics[width=\columnwidth]{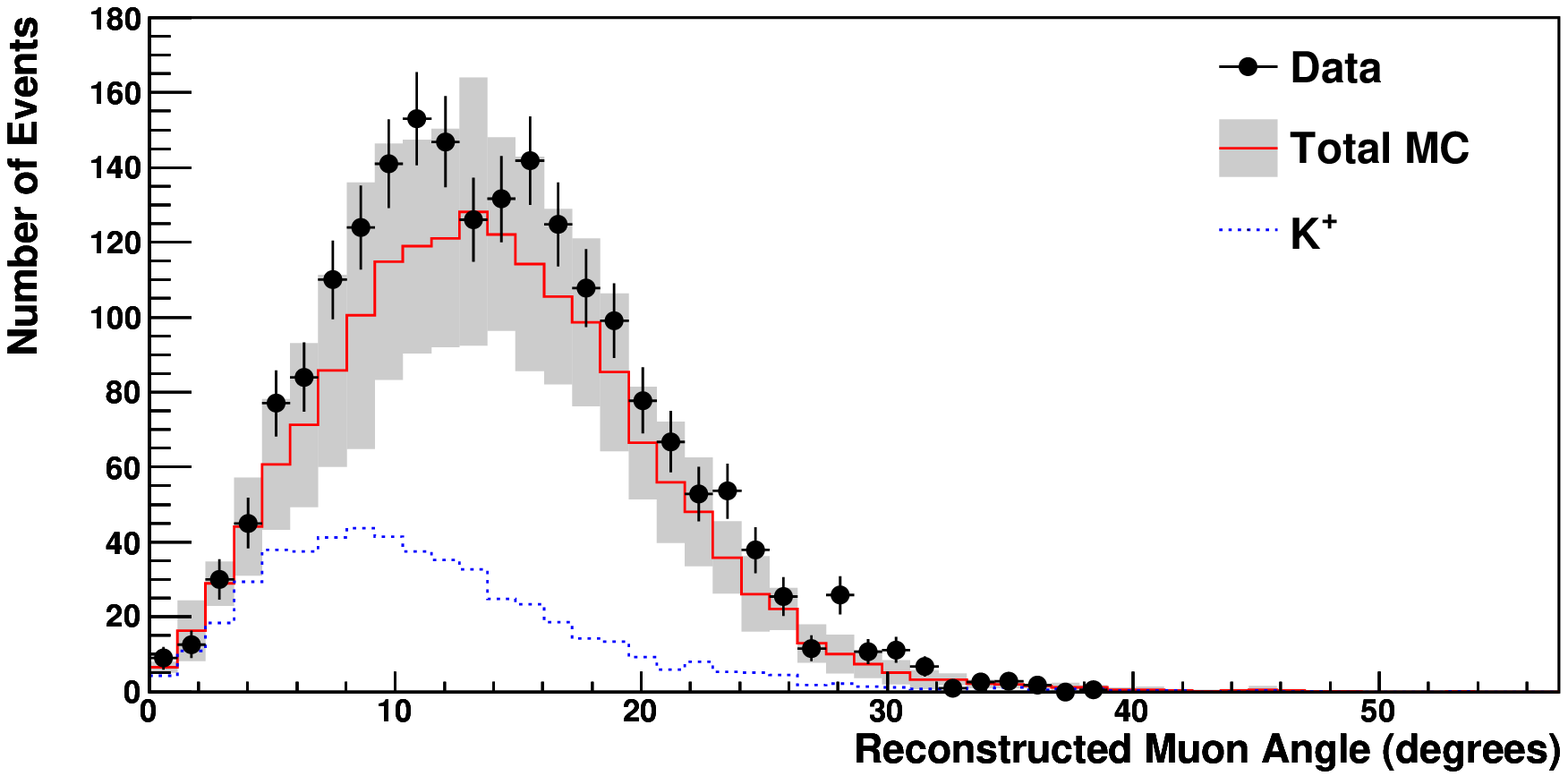}}
\subfigure[~2-Track Sample]{\includegraphics[width=\columnwidth]{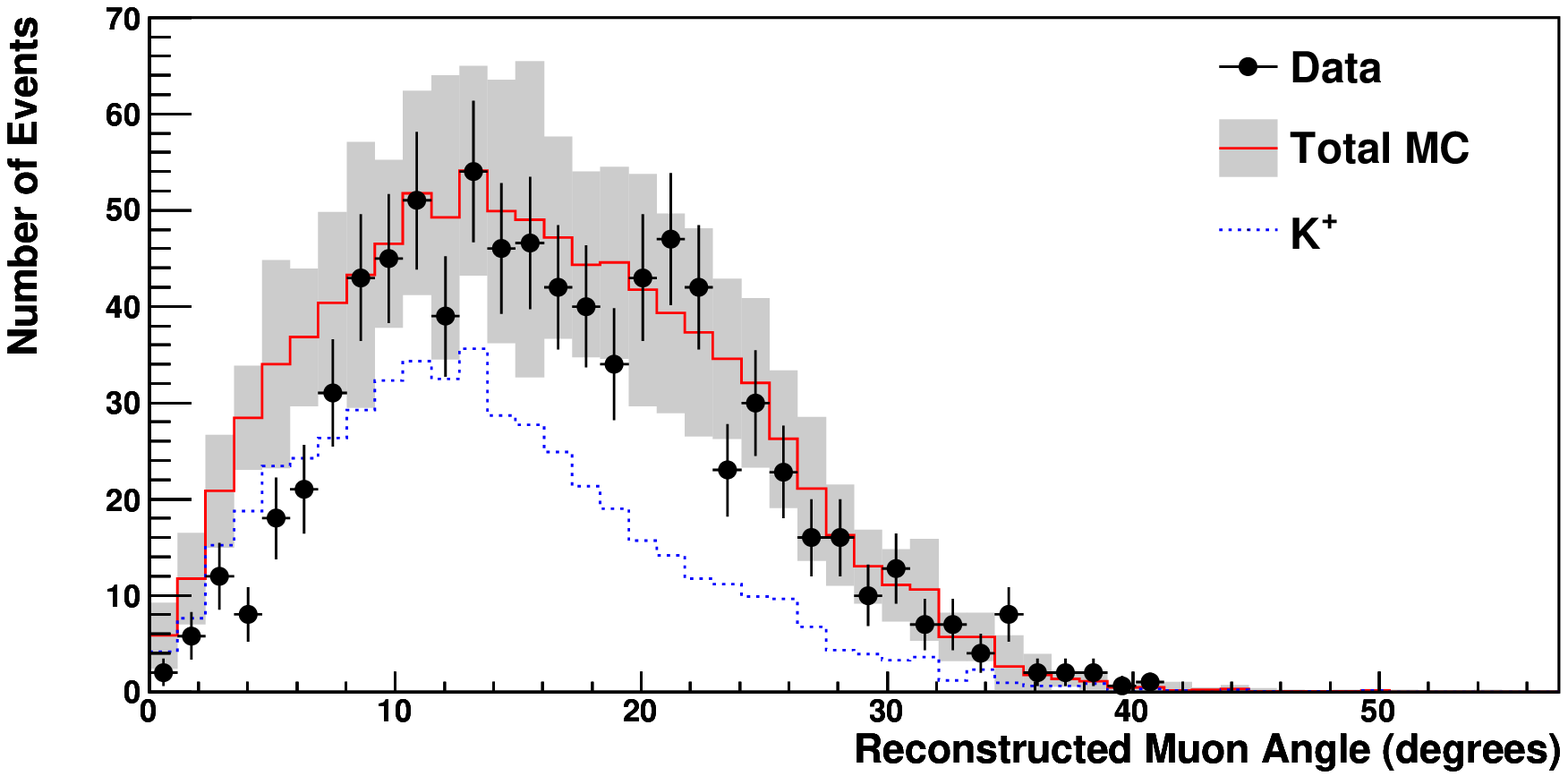}}
\subfigure[~3-Track Sample]{\includegraphics[width=\columnwidth]{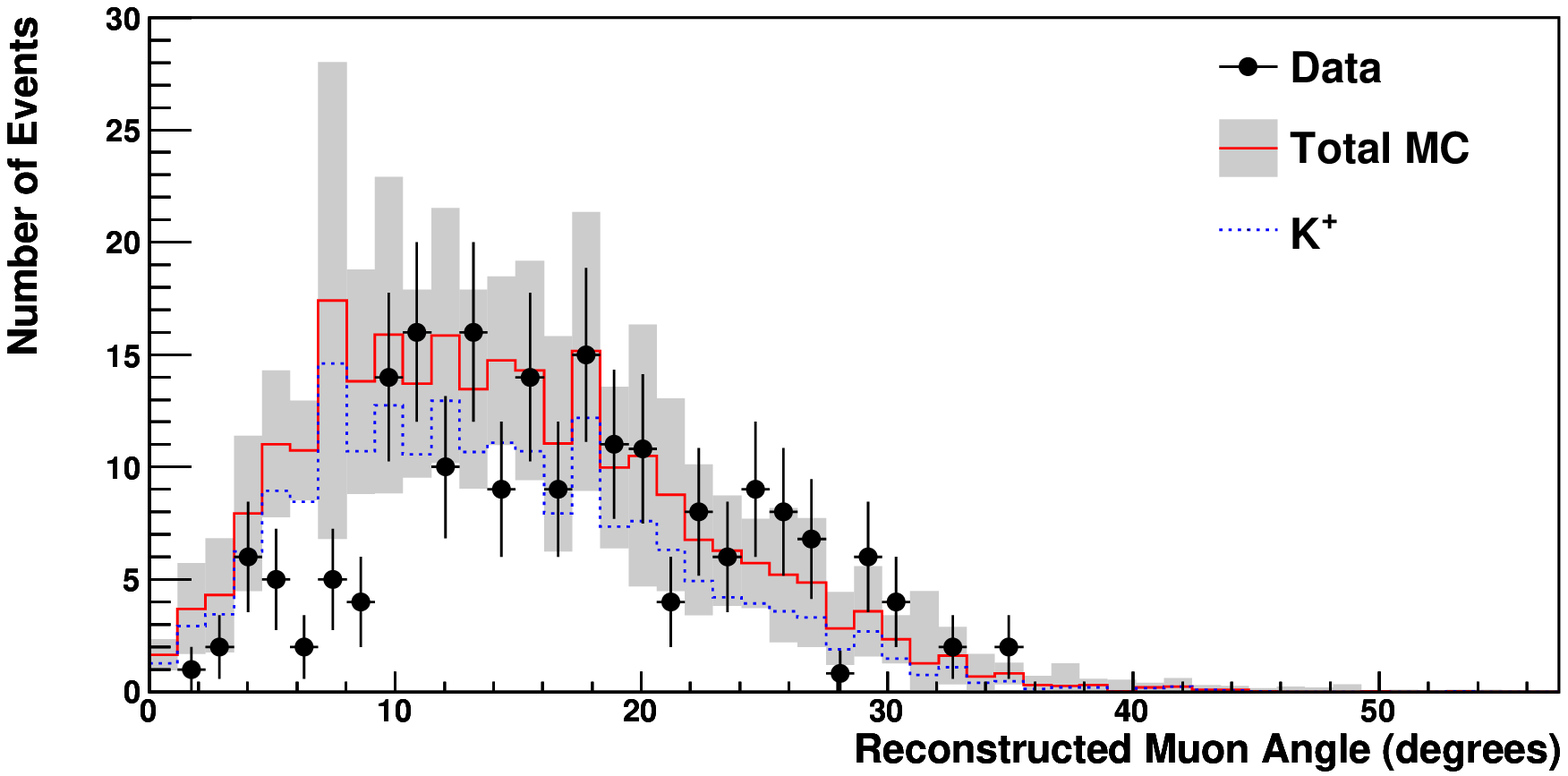}}
\caption{Reconstructed muon angle for the SciBar 1-Track, 2-Track, and 3-Track samples in the neutrino mode analysis. The background contributions from $K^{-}$ and $K^{0}_{L}$ are very small but included in the MC histogram. The grey area represents the total systematic uncertainty in the MC.}
\label{fig:selected_samples}
\end{center}
\end{figure}

After all the previously mentioned selection cuts, the resulting data and MC samples are further divided into three separate samples based on the number of SciBar reconstructed tracks in each event. The number of SciBar reconstructed tracks for data and MC (along with its different contributions) can be seen in Fig.~\ref{fig:selectedsample_ntracks}. The SciBar 1-Track sample contains mostly (78\%) CCQE interactions, the SciBar 2-Track sample contains both CCQE(46\%) and CC1$\pi$ (42\%) and the SciBar 3-Track sample contains mostly CC1$\pi$(62\%) events together with QE and DIS interactions. The reconstructed muon angle distributions for the SciBar 1, 2 and 3-Track samples are shown in Fig.~\ref{fig:selected_samples}. Fig.~\ref{fig:selected_samples} shows an excess in the MC events with respect to the data for low angle muon reconstructed tracks which is largely attributed to the difficulty in correctly modeling nuclear final state interactions. We believe that this MC/data discrepancy at low reconstructed muon angle is due to modeling of proton emission; the effect is covered by the uncertainty included for the modeling in Sec.~\ref{sec:sysuncertain}.

Tab.~\ref{tab:summary_fitted} shows the number of events coming from $K^+$, $\pi^+$, data, and MC (NUANCE) along with efficiencies and purities (defined in Eqs.~\ref{eq:numu_efficiency},~\ref{eq:numu_purity}) for the three selected samples.

\begin{table}[htbn!]
\begin{center}
\caption{Number of events for data and for NUANCE MC after selection cuts for the neutrino mode analysis. MRD indicates the SciBar-MRD sample, and MRD pen. stands for SciBar-MRD penetrated sample. The cosmic backgrounds have been subtracted from the data. The first column contains the number of predicted events with a $\nu_\mu$ coming from $K^+$ while the second column contains the number of predicted events with a $\nu_\mu$ coming from $\pi^{+}$. The third column and fourth column represent the number of events in data and MC and the last two columns are the efficiency ($\epsilon(K^{+})$) and purity ($\pi(K^{+})$) for $\nu_\mu$ from $K^{+}$ in MC. The total event prediction from simulation is labeled as MC.}
\vspace{0.5cm}
\begin{tabular}{c|c|c|c|c|c|c}\hline\hline
Event Sel.                                        & $K^{+}$$\nu_\mu$       & $\pi^{+}$$\nu_\mu$   & Data     & MC    & $\epsilon(K^{+})$        & $\pi(K^{+})$ \\\hline
MRD                                               & 2,867                             & 18,173                      & 27,049  & 22,142       & 58\%         & 13\%    \\
MRD pen.                                       & 1,508                             & 1,700                       & 3,365     & 3,286         & 31\%         & 46\%    \\
Single $\mu$                                   & 1,313                             & 1,666                       & 3,188      & 3,053         & 27\%         & 43\%    \\
1-Trk                                              & 509                                & 1,159                       & 2,050      & 1,723         & 10\%         & 30\%    \\
2-Trk                                              & 497                                & 438                         & 834          & 950           & 10\%         & 52\%    \\
3-Trk                                              & 189                                & 57                          & 206           & 250           & 4\%          & 76\%    \\
\hline\hline
\end{tabular}
\label{tab:summary_fitted}
\end{center}
\end{table}

\begin{figure}[htbn!]
\begin{center}
\subfigure[~1-Track Sample]{\includegraphics[width=\columnwidth]{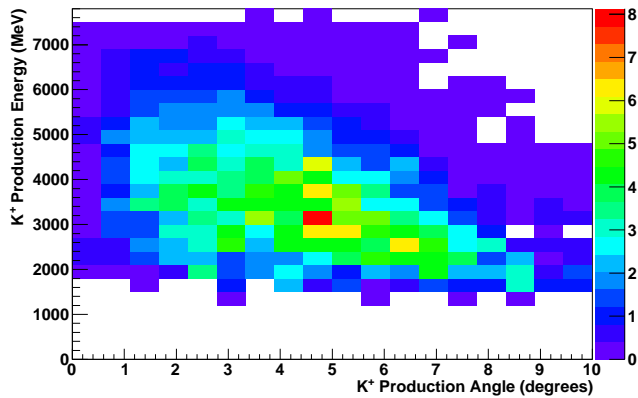}}
\subfigure[~2-Track Sample]{\includegraphics[width=\columnwidth]{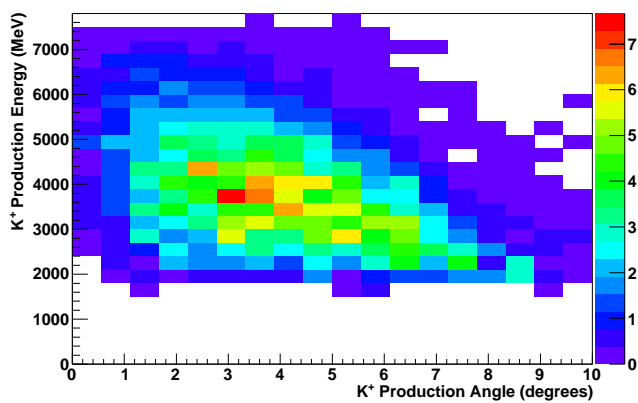}}
\subfigure[~3-Track Sample]{\includegraphics[width=\columnwidth]{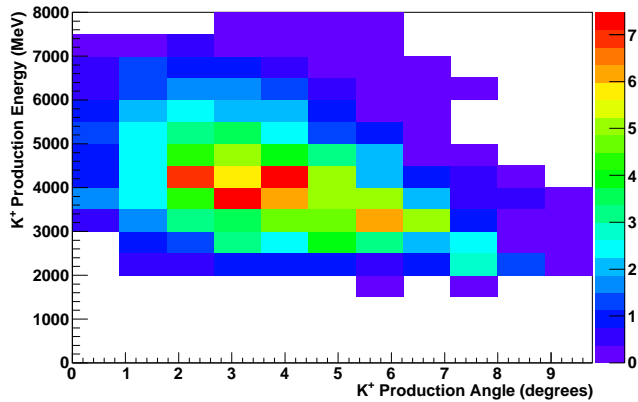}}
\caption{Two-dimensional plots of the energy vs angle relative to the proton beam axis for the $K^+$ selected events in the SciBar 1-Track, 2-Track, and 3-Track MC samples for the neutrino mode analysis.}
\label{fig:neutrinomode2Dhistos}
\end{center}
\end{figure}

We define the efficiency and purity using MC simulated events as:
\begin{eqnarray}
&&\epsilon_{K^{+}} = \frac{\#~selected~events}{\#~generated~K^{+}}, \label{eq:numu_efficiency} \\
&&\pi_{K^+} = \frac{\#~true~K^{+}~in~selected~sample}{\#~selected~events}, \label{eq:numu_purity}
\end{eqnarray}
The mean energy and mean angle (with respect to proton beam direction) of the selected $K^+$ and the mean energy for $\nu_\mu$ from the selected $K^+$ in each of the three samples is summarized in Tab.~\ref{tab:numu_energy}. Fig.~\ref{fig:neutrinomode2Dhistos} shows the 2-dimensional distribution of $K^+$ production angle relative to proton beam axis versus true $K^+$ production energy for predicted $K^+$ events selected in the SciBar 1-Track, 2-Track and 3-Track MC samples for the neutrino mode analysis.
\begin{table}[htbn!]
\begin{center}
\caption{NUANCE MC predicted mean energy and mean angle (with respect to the proton beam direction) for the selected $K^+$ samples and predicted MC energy for the selected $\nu_\mu$ from $K^+$ for the three SciBar samples in neutrino mode running. Errors correspond to the error on the mean values. RMS of the relative distributions is reported between squared parenthesis.}
\vspace{0.5cm}
\begin{tabular}{c|c|c|c}\hline\hline
         & $E_{K^+}$[RMS](GeV) & $\theta_{K^+}$[RMS](deg)  & $E_{\nu_\mu}$[RMS](GeV) \\ \hline
1-Trk  & 3.6$\pm$0.1[1.2]     & 4.3$\pm$0.1[2.1]             & 3.0$\pm$0.1[1.0]   \\ \hline
2-Trk  & 3.8$\pm$0.1[1.2]     & 4.1$\pm$0.1[2.0]             & 3.3$\pm$0.1[1.0]   \\ \hline
3-Trk  & 4.1$\pm$0.1[1.1]     & 3.9$\pm$0.1[1.9]             & 3.5$\pm$0.1[0.9]   \\ \hline\hline
\end{tabular}
\label{tab:numu_energy}
\end{center}
\end{table}

The background from $K^{-}$ and $K^{0}_{L}$ decays are very small. The expected number of $K^{-}$ and $K^{0}_{L}$ events predicted for all three samples combined are 5 and 3 events, respectively.


 \section{Antineutrino Mode Analysis}\label{sec:antinumu_analyisis}

The dataset used in this analysis consists of antineutrino-mode events collected between June 2007 and August 2008 corresponding to a total of $1.51 \times 10^{20}$ POT, after all data quality cuts. The data collection efficiency was also 95{\%} in this case.

The majority of events in antineutrino mode running are $\bar{\nu}_\mu$ from $\pi^{-}$ decay, with a significant $\nu_\mu$ contribution coming from the decay of positively charged $\pi^{+}$ and $K^{+}$. At higher energies (above 2 GeV), the positively charged ``wrong-sign" particles, mainly $\pi^+$ and $K^+$, are strongly boosted in the forward direction, traveling downstream nearly parallel to the beam axis direction, such that they are not defocused by the magnetic horn. These particles create a 40\% background of wrong sign neutrino interactions in the antineutrino mode run.
In this analysis, we select a sample of high energy $\nu_\mu$ coming from $K^{+}$. The analysis strategy is very similar to what is described in Sec.~\ref{sec:muon_neutrino_analysis} with the difference that the background composition in the selected sample is different. Background events result from high energy $\nu_\mu$ from $\pi^{+}$ produced in the initial p-Be interaction, and the decay of $\pi^{-}$ (also produced in the initial p-Be interaction) that generate high energy $\bar{\nu}_\mu$. Both the $\nu_\mu$ and $\bar{\nu}_\mu$ result in high energy muons (with negative and positive charge respectively) that penetrate the SciBar, EC and MRD detectors. Positive and negative muons are indistinguishable in SciBooNE due to the lack of magnetic field in our detector system.

\begin{figure}[htbn!]
\begin{center}
\includegraphics[width=\columnwidth]{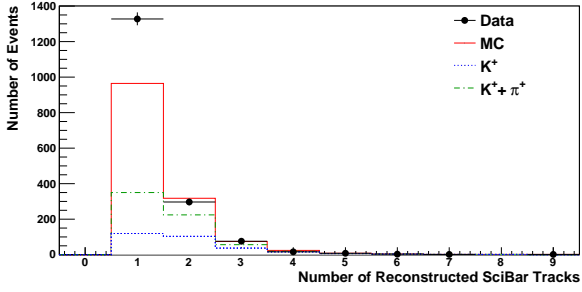}
\caption{Number of SciBar reconstructed tracks for the selected event sample in antineutrino mode for data and NUANCE MC. MC histogram includes all events and $K^+$ shows the component of $\nu_\mu$ from $K^+$ and $K^{+} + {\pi^+}$ shows the component of $\nu_\mu$ from $K^+$ and $\pi^+$.}
\label{fig:antinumu_selectedsample_ntracks}
\end{center}
\end{figure}

\begin{figure}[htbn!]
\begin{center}
\subfigure[~1-Track Sample]{\includegraphics[width=\columnwidth]{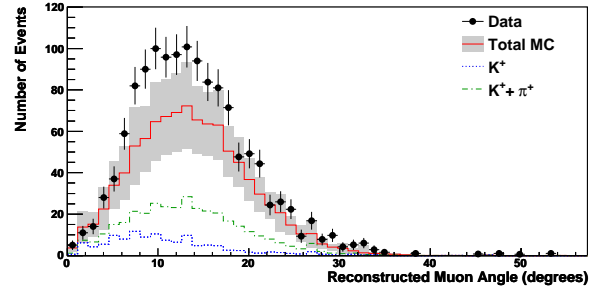}}
\subfigure[~2-Track Sample]{\includegraphics[width=\columnwidth]{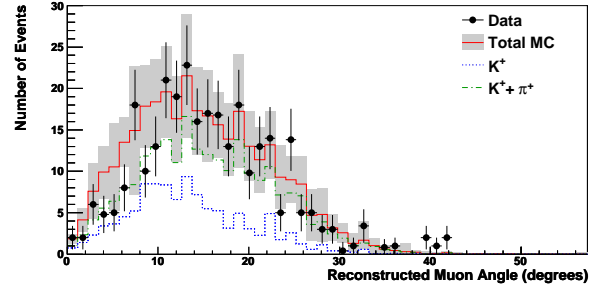}}
\subfigure[~3-Track Sample]{\includegraphics[width=\columnwidth]{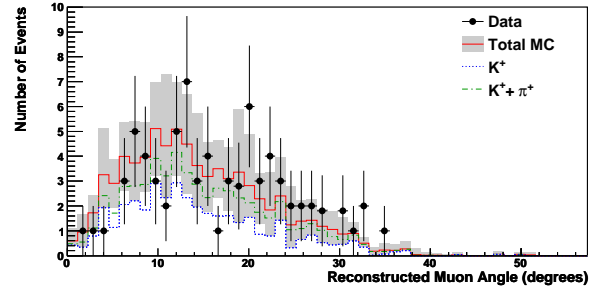}}
\caption{Reconstructed muon angle for the SciBar 1-Track, 2-Track, and 3-Track sample in the antineutrino mode analysis. The background contributions from $K^{-}$ and $K^{0}_{L}$ are small but included in the MC histogram. The grey area represents the total systematic uncertainty in the MC.}
\label{fig:antinumu_selectedsample}
\end{center}
\end{figure}

As in the neutrino mode analysis, events passing the base selection cuts are further divided into three samples based on whether the events contain 1, 2 or 3 SciBar reconstructed tracks. The number of SciBar reconstructed tracks for data and MC (along with the parent) can be seen in Fig.~\ref{fig:antinumu_selectedsample_ntracks}. The SciBar 1-Track sample contains mostly (81\%) charged current quasi-elastic interaction events. The SciBar 2-Track sample is evenly split between charged current quasi-elastic (40\%) and charged current resonant pion interactions (44\%), with a tiny contribution from charged current multi-$\pi$/DIS interactions (6\%). The SciBar 3-Track sample contains mostly charged current resonant pion interactions (61\%), with small contributions from charged current quasi-elastic (12\%) and charged current multi-$\pi$/DIS interactions (21\%). The reconstructed muon angle distributions for the SciBar 1, 2 and 3-Track samples are shown in Fig.~\ref{fig:antinumu_selectedsample}. In this case the $\nu_\mu$ events coming from $K^+$ have high energies, peaking at smaller angles, while the lower energy $\pi^{+}$ and $\pi^{-}$ background distributions are spread more evenly across a broader range of angles.

In this analysis, we also found a slight disagreement between data and Monte Carlo at low angle regions for the SciBar 2 and 3-Track samples for the reconstructed muon angle (see Fig.~\ref{fig:antinumu_selectedsample}). This disagreement is caused, as in the $\nu_\mu$ analysis, by the imperfect modeling of nuclear and intra-nuclear interactions in our neutrino interaction simulation program.

Tab.~\ref{tab:antinumu_selected_events} summarizes the selected number of events in Data and MC (NUANCE) for the three selected samples, together with the efficiency and purity of $K^+$ (defined in Eqs.~\ref{eq:numu_efficiency},~\ref{eq:numu_purity}) .

\begin{table}[htbn!]
\begin{center}
\caption{Number of events for data and for NUANCE MC after selection cuts for the antineutrino mode analysis. MRD indicates the SciBar-MRD sample, and MRD pen. stands for SciBar-MRD penetrated sample. The cosmic backgrounds have been subtracted from the data. The first column contains the number of predicted events with $\nu_\mu$ coming from $K^+$, the second column contains the number of predicted events with $\nu_\mu$ coming from $\pi^+$, and the third column contains the number of predicted events with $\bar{\nu}_\mu$ coming from $\pi^-$. Fourth column and fifth column represent the number of events in data and MC and the last two columns are the efficiency ($\epsilon(K^{+})$) and purity ($\pi(K^{+})$) for $\nu_\mu$ from $K^+$ in MC. The total event prediction from simulation is labeled as MC.}
\vspace{0.5cm}
\begin{tabular}{c|c|c|c|c|c|c|c}\hline\hline
Event Sel.              & $K^{+}$$\nu_\mu$      & $\pi^{+}$$\nu_\mu$   & $\pi^{-}$$\bar{\nu}_\mu$     & Data  & MC        & $\epsilon(K^{+})$     & $\pi(K^{+})$ \\\hline
MRD                     & 705                   & 2,287                & 6,167                    & 11,528   & 9,499      & 52\%                & 7\%   \\
MRD pen.             & 326                   & 385                  & 698                         & 1,790    & 1,452       & 24\%                & 22\%    \\
Single $\mu$         & 283                   & 375                  & 691                         & 1,728    & 1,389       & 21\%                & 20\%    \\
1-Trk                    & 119                   & 230                  & 589                         & 1328     & 965          & 9\%                 & 12\%    \\
2-Trk                    & 103                   & 121                  & 83                          & 296        & 317           & 8\%                 & 32\%    \\
3-Trk                    & 36                    & 19                   & 15                            & 75          & 74            & 3\%                 & 49\%    \\\hline\hline
\end{tabular}
\label{tab:antinumu_selected_events}
\end{center}
\end{table}

\begin{figure}[htbn!]
\begin{center}
\subfigure[~1-Track Sample]{\includegraphics[width=\columnwidth]{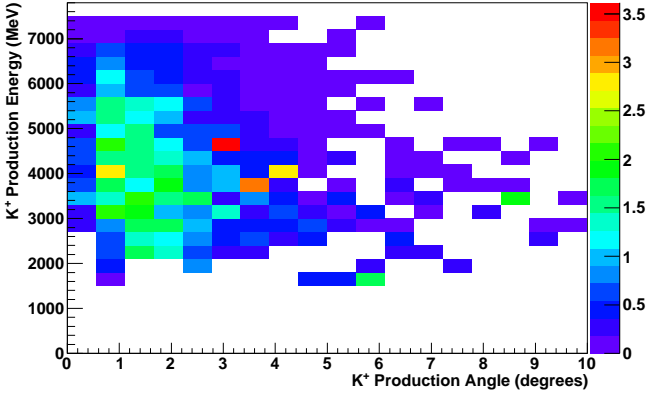}}
\subfigure[~2-Track Sample]{\includegraphics[width=\columnwidth]{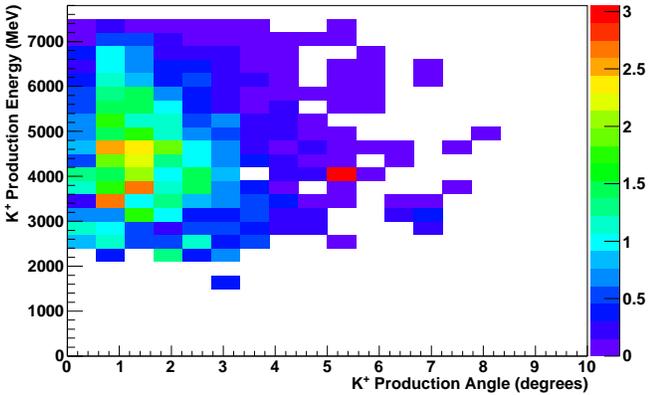}}
\subfigure[~3-Track Sample]{\includegraphics[width=\columnwidth]{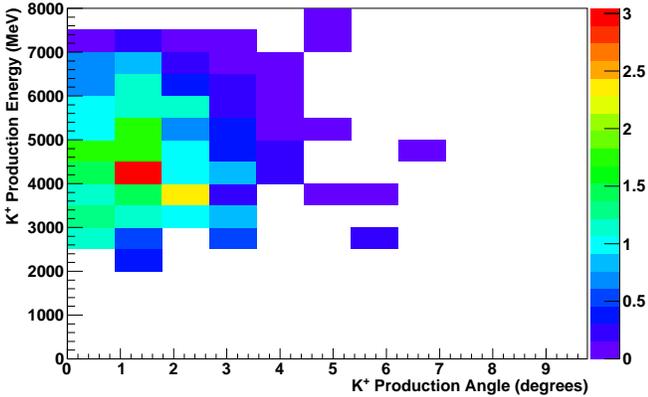}}
\caption{Two-dimensional plots of the energy vs angle relative to the proton beam axis for the $K^+$ selected events in the SciBar 1-Track, 2-Track, and 3-Track MC samples for the antineutrino mode analysis. The $K^+$ are much more forward than in neutrino mode (Fig.~\ref{fig:neutrinomode2Dhistos}) otherwise they would have been swept out by the horn, this can be seen also in the angular distribution difference between Tab.~\ref{tab:numu_energy} and Tab.~\ref{tab:antinumu_energy}.}
\label{fig:antineutrinomode2Dhistos}
\end{center}
\end{figure}

The mean energy and mean angle (with respect to the proton beam direction) for the selected $K^+$ and the mean energy for $\nu_\mu$ from the selected $K^+$ in each of the three samples is summarized in Tab.~\ref{tab:antinumu_energy}. Fig.~\ref{fig:antineutrinomode2Dhistos} shows the 2-dimensional distribution of $K^+$ production angle relative to proton beam axis versus true $K^+$ production energy for predicted $K^+$ events selected in the SciBar 1-Track, 2-Track and 3-Track MC samples for the antineutrino mode analysis. As can
be seen in Fig.~\ref{fig:antineutrinomode2Dhistos}, the $K^+$ are much more forward than in neutrino mode (Fig.~\ref{fig:neutrinomode2Dhistos}) otherwise they would have been swept out by the horn.

\begin{table}[htbn!]
\begin{center}
\caption{NUANCE MC predicted mean energy and mean angle (with respect to the proton beam direction) for the selected $K^+$ samples and predicted MC energy for the selected $\nu_\mu$ from $K^+$ for the three SciBar samples in antineutrino mode running. Errors are the error on the mean values. RMS of the relative distributions is reported between squared parenthesis.}
\vspace{0.5cm}
\begin{tabular}{c|c|c|c}\hline\hline
          & $E_{K^+}$[RMS](GeV) & $\theta_{K^+}$[RMS](deg) & $E_{\nu}$[RMS](GeV) \\ \hline
1-Trk   & 4.1$\pm$0.1[1.2]     & 2.4$\pm$0.2[1.8]            & 3.1$\pm$0.1[1.1] \\ \hline
2-Trk   & 4.4$\pm$0.1[1.2]     & 1.9$\pm$0.1[1.4]            & 3.4$\pm$0.1[1.0] \\ \hline
3-Trk   & 4.6$\pm$0.2[1.1]     & 1.6$\pm$0.2[1.0]            & 3.6$\pm$0.2[1.0] \\ \hline\hline
\end{tabular}
\label{tab:antinumu_energy}
\end{center}
\end{table}

\begin{figure}[htbn!]
\begin{center}
\includegraphics[width=\columnwidth]{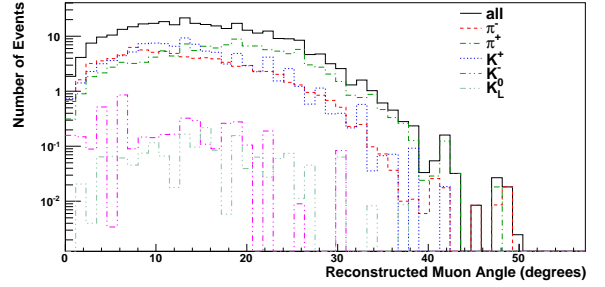}
\caption{The various meson contributions that decay into $\nu_\mu$ and $\bar{\nu}_\mu$ as a function of reconstructed muon angle in the 2-Track sample in antineutrino mode.}
\label{fig:anti_kminus_kzero_muon_angle}
\end{center}
\end{figure}

The number of neutrinos coming from $K^{-}$ and $K^{0}_{L}$ decays are small: the number of expected $K^{-}$ and $K^{0}_{L}$ events predicted for all three samples combined are 25 and 5 events, respectively. Fig.~\ref{fig:anti_kminus_kzero_muon_angle} shows the reconstructed muon angle for the 2-Track sample with the $K^{-}$ and $K^{0}_{L}$ contributions. Interactions from cosmic backgrounds, neutrino interactions in the EC and MRD, and events with neutrino interactions in the material surrounding the SciBar detector make a negligible contribution in this analysis as well. For all three samples combined, 19 events are estimated from cosmic backgrounds, 1 event is estimated from backscattering neutrino events in EC and MRD, and 3 events are estimated from neutrino interactions outside the SciBooNE detector.

\section{Systematic Uncertainties}
\label{sec:sysuncertain}

Systematic uncertainties are included by using a covariance matrix that includes the correlated and uncorrelated errors among the muon angle bins in the SciBar 1, 2, 3-Track samples. The covariance matrix is determined by calculating the correlated event changes for a given systematic uncertainty and then combining these (the different systematic uncertainties are assumed to be uncorrelated) to form the matrix. Statistical errors are also included as uncorrelated terms in the diagonal elements.

\subsection{Neutrino Beam Uncertainties}

The uncertainties in the delivery of the primary proton beam to the beryllium target and the primary beam optics, secondary hadron production in proton-beryllium interactions, hadronic interactions in the target and horn and the horn magnetic field model are considered. The change in the neutrino beam spectrum due to these uncertainties is calculated by drawing random parameter vectors (all the beam systematics are varied within their uncertainties when drawing random parameter vectors) and weighting each event by a factor corresponding to the variation of the yield of the parent meson with the given momentum and angle.

To evaluate the neutrino beam uncertainties for both the neutrino and antineutrino mode running, a thousand random parameter vectors are generated, resulting in a thousand neutrino beam flux predictions and a thousand individual weights for each MC event. The thousand individual sets of weights for the MC simulation are passed through each analysis resulting in a thousand individual outcomes for each analysis. The correlated event changes associated with the thousand individual outcomes are used to form the error matrix for the beam systematic uncertainties.

For $K^-$ production error, which is not accounted for in the previous neutrino beam uncertainty weights, a conservative 100\% production uncertainty is applied.

\subsection{Detector Uncertainties}

\subsubsection{PMT Crosstalk and Resolution}

The crosstalk of the MA-PMT was measured to be 3.15\% for adjacent channels, with an absolute error of 0.4\%~\cite{Hiraide:2008eu}. The single photoelectron resolution of the MA-PMT is set to 50\% in the simulation, and the absolute error is estimated to be $\pm 20$\%. Several complete MC simulation sets, with the crosstalk level and single photoelectron resolution separately varied within their uncertainties, are prepared and the changes in the final results using the varied MC simulation sets are taken as the systematic uncertainties.

\subsubsection{Scintillator Quenching}

Birk's constant for the SciBar scintillator was measured to be $0.0208 \pm 0.0023$ cm/MeV~\cite{Hiraide:2008eu} and is varied within the measurement to evaluate the systematic error.

\subsubsection{Hit Threshold}

The conversion factors from the ADC counts to the photoelectron were measured for all 14,336 MA-PMT channels in SciBar. The measurement uncertainty was at the 20\% level. Since the number of photoelectrons (2.5~p.e.) for the SciBar hit threshold was used, the hit threshold is varied by $\pm$20\% to evaluate the systematic error for SciBar track reconstruction.

\subsubsection{TDC Dead time}

The TDC dead time is set to 55 ns in the MC simulation, with the error estimated to be $\pm$20~ns~\cite{Hiraide:2009zz}. A MC simulation set with variations in the TDC dead time was prepared to determine the systematic error.

\subsection{Cross section and nuclear model uncertainties}

\subsubsection{CC Quasi Elastic Scattering Cross Section}

A systematic uncertainty of $\pm$0.234~GeV is assigned to the $M_A^{QE}$-C to span the difference between the value used and the value from the global fit to historical data~\cite{Bodek:2007vi}. The difference between $\kappa = 1.000$ and $\kappa = 1.022$ is also assigned as a systematic uncertainty and added in quadrature to the $M_A^{QE}$-C error.  $M_A^{QE}$-H error gives a negligible contribution.

\subsubsection{CC Resonant Pion Production Cross Section}

From a previous K2K measurement~\cite{Rodriguez:2008eaa}, the uncertainty in the resonant pion scattering cross section is estimated to be $\pm$20\%. An additional uncertainty is assigned to account for the observed $Q^2$ disagreement between the SciBooNE CC 1$\pi$-enriched data samples and the MC~\cite{Nakajima:2010fp}. The uncertainty is evaluated by re-weighting CC resonant pion events as a function of true $Q^2$ such that they match the observed distribution in the SciBooNE data. Resonance decays leading to multi-pion final states are also included in the model and are simulated assuming $M_A^{N\pi}=1.30$ GeV/$c^2$. This value of $M_A^{N\pi}$ is chosen strictly to ensure that the total CC cross section prediction reproduces previous experimental data.

\subsubsection{Multi Pion Production and Deep Inelastic Scattering Cross Section}

The uncertainties for multi pion and deep inelastic scattering (DIS) cross sections are set respectively at $\pm 40\%$. The 40\% multi-pion and DIS uncertainties come from a comparison of the NUANCE predictions to the existing multi-pion data in~\cite{Kitagaki:1986ct,Day}.

\subsubsection{Pion Interaction in the Initial Target Nucleus}

For pions produced by neutrino interactions, uncertainties on the cross sections for pion absorption, pion inelastic scattering, and pion charge exchange in the nucleus are considered. The values of absorption, inelastic scattering, and charge exchange for pions are varied by $\pm$30\% in the MC independently while keeping the total number of neutrino events with pion processes fixed to determine the systematic uncertainties of each process.

\subsubsection{Proton Emission}

A systematic uncertainty associated with the emission of a proton following pion absorption in the nucleus is also included. The proton emission following a pion absorption alters both the number of reconstructed tracks and the event kinematics. Turning off this nuclear process in the MC results in better agreement between the data and the MC, thus, indicating that the modeling may be inaccurate. We have estimated the uncertainty associated with proton emission by calculating the difference for the MC with this process turned on and off. The uncertainty is calculated for both NUANCE and NEUT, with NEUT giving the larger systematic uncertainty, so the uncertainty is conservatively estimated using NEUT for both MCs.


\section{Fits to determine the $K^+$ Production and Rate Normalization}\label{sec:covariancefit}

The final state reconstructed track multiplicity and the muon angular distributions for the selected high energy events are different for neutrinos produced by pion and kaon decays as presented in Sec.~\ref{sec:muon_neutrino_analysis} and in Sec.~\ref{sec:antinumu_analyisis}. Therefore, we fit these distributions in order to isolate neutrinos from kaon decays and minimize the following $\chi^2$ function:

\begin{eqnarray}
\nonumber && {\chi}^2\: =\: {\chi}^2_{\nu\:}\: +\: {\chi}^2_{\bar{\nu}\:} =\\
\nonumber && \sum_{i,j}^{N} (N_i^{obs} - N_i^{pred}){(V^{\nu}_{stat} + V^{\nu}_{sys})}_{ij}^{-1}(N_j^{obs} - N_j^{pred}) +\\
&& \sum_{p,q}^{M} (M_p^{obs} - M_p^{pred}){(V^{\bar{\nu}}_{stat} + V^{\bar{\nu}}_{sys})}_{pq}^{-1}(M_q^{obs} - M_q^{pred}).
\label{eq:chi2}
\end{eqnarray}

Cross section values (CCQE and CC1$\pi$ for $\nu$ and $\bar{\nu}$) and uncertainties are included in the fit as described respectively in Appendix~\ref{app:xsect-fit} and in Sec.~\ref{sec:sysuncertain}.

Two separate $K^{+}$ normalizations with respect to the beam MC predictions (Eq.~\ref{eq:central_value_doublediff}) are extracted from the $K^{+}$ selected samples, a $K^{+}_{prod}$, the $K^{+}$ production flux at the beam target, and a $K^{+}_{rate}$, the $K^{+}_{prod} \times$ neutrino cross-sections.
These two types of determinations are needed since the $K^{+}_{prod}$ normalization can be used to determine a measured $K^+$ production cross section and the $K^{+}_{rate}$ normalization can be used by BNB experiments as better estimates of the $K^+$ production rate in their beam.

The ${\chi}^2$ function in Eq.~\ref{eq:chi2} contains two terms: the former ${\chi}^2_{\nu\: }$ term is associated with events for neutrino mode running and the latter ${\chi}^2_{\bar{\nu}\:}$ term is associated with events for antineutrino mode running. The two $\chi^2$ functions are assumed to be effectively uncorrelated since the cross section uncertainties for the antineutrino mode data are small compared to the statistical and background uncertainties. In the neutrino mode running, all three samples (SciBar 1, 2, 3-Track) are used simultaneously in the fit including their correlated bin-to-bin uncertainties. Only bins with 10 or more events are included in the ${\chi}^2$. N and M are the number of bins used in the three reconstructed angle distributions in neutrino and antineutrino mode, respectively. $N_{i(j)}^{obs}$ and $N_{i(j)}^{pred}$ are the numbers of observed and predicted events in the $i(j)$-th angle bin for the neutrino mode analysis. $M_{p(q)}^{obs}$ and $M_{p(q)}^{pred}$ are the same quantities for the antineutrino in the $p(q)$-th angle bin.

For the $K^+$ production analysis, the functions that describe the number of predicted events $N_{i(j)}^{pred}$ and $M_{p(q)}^{pred}$ are given by Eq.~\ref{eq:pred_for_production_i} for neutrinos and Eq.~\ref{eq:pred_for_production_antinu} for antineutrinos in Appendix~\ref{app:xsect-fit}. $(V^{\nu}_{sys})_{ij}$ and $(V^{\bar{\nu}}_{sys})_{pq}$ are the elements of the covariance matrix for neutrino and antineutrino mode for each of the systematic uncertainties described in Sec.~\ref{sec:sysuncertain}. $V^{\nu}_{stat}$ ($V^{\bar{\nu}}_{stat}$) represents the statistical error in neutrino mode running (antineutrino mode running). An initial neutrino mode and antineutrino mode combined ${\chi}^2$ minimization is performed to determine the best cross-section normalization factors for both $\nu_\mu$ and $\bar{\nu}_\mu$ as described in Appendix~\ref{app:xsect-fit}. Pull terms on the cross-sections normalization factors are added to keep the minimization physical. After the initial combined ${\chi}^2$ minimization, the cross-section weights are fixed in $N_{i}^{pred}$ and $M_{p}^{pred}$ at their minimized values and the pull terms are removed from the ${\chi}^2$ to evaluate the total (statistical+systematical) uncertainty on the $K^+$ production or rate. These cross-section weights are initially minimized to allow for better agreement between data and MC in the plots and do not affect either $K^{+}$ production or rate weights because the large uncertainties on the neutrino cross-section values are already taken into account in the covariance matrix.

\begin{table}[htbp!]
  \begin{center}
  \caption{Summary of the fit results for the cross-section normalization factors as described in Appendix \ref{app:xsect-fit} with respect to NUANCE predictions. The top four values are cross section normalization values for $\nu_\mu$ and $\bar{\nu}_\mu$ coming from $\pi^+$ and $K^+$ while the last two are normalization factors for $\bar{\nu}_{\mu}$ coming from $\pi^-$.}
  \vspace{0.5cm}
  \begin{tabular}{c|c|c}\hline\hline
    $\pi^+$/$K^+$ & & Fit Value \\\hline
    & CCQE in $\nu$ mode         & 1.17$\pm$0.14 \\
    & CCQE in ${\bar{\nu}}$ mode & 1.07$\pm$0.25 \\
    & CC1$\pi$ in $\nu$ mode       & 0.89$\pm$0.25 \\
    & CC1$\pi$ in ${\bar{\nu}}$ mode & 0.91$\pm$0.26 \\\hline
    $\pi^-$ && \\\hline
    & CCQE in ${\bar{\nu}}$ mode & 1.50$\pm$0.21 \\
    & CC1$\pi$ in ${\bar{\nu}}$ mode & 1.49$\pm$0.29 \\
    \hline\hline
  \end{tabular}
  \label{tab:nuance_fit_results_xsect}
  \end{center}
\end{table}

The $K^{+}$ rate is measured by minimizing the same $\chi^2$ function as described in Eq.~\ref{eq:chi2} but using Eq.~\ref{eq:pred_for_rate_i} and Eq.~\ref{eq:pred_for_rate_antinu} respectively for neutrino and antineutrino to predict the number of events. The covariance matrices used for the $K^+$ rate
measurement does not include the cross section uncertainties for the $\nu_{\mu}$ from $K^+$ in contrast to the matrices used for the flux measurement. A summary of the cross section normalizations is presented in Tab.~\ref{tab:nuance_fit_results_xsect}. Many of these values are consistent with low energy precision cross-section measurements from the MiniBooNE experiment~\cite{AguilarArevalo:2008yp,teppei:2007ru,AguilarArevalo:2010bm,AguilarArevalo:2011sz} though the two sets of cross-section values are measured at different energies. The MiniBooNE collaboration measures cross-sections at neutrino energies less than 2 GeV while the cross-section values listed in Tab.~\ref{tab:nuance_fit_results_xsect} are at neutrino energies greater than 3 GeV (as could be seen from Tab.~\ref{tab:numu_energy} and Tab.~\ref{tab:antinumu_energy}).

\begin{figure}[htbp!]
\begin{center}
\subfigure[~Neutrino Sample]{\includegraphics[width=\columnwidth]{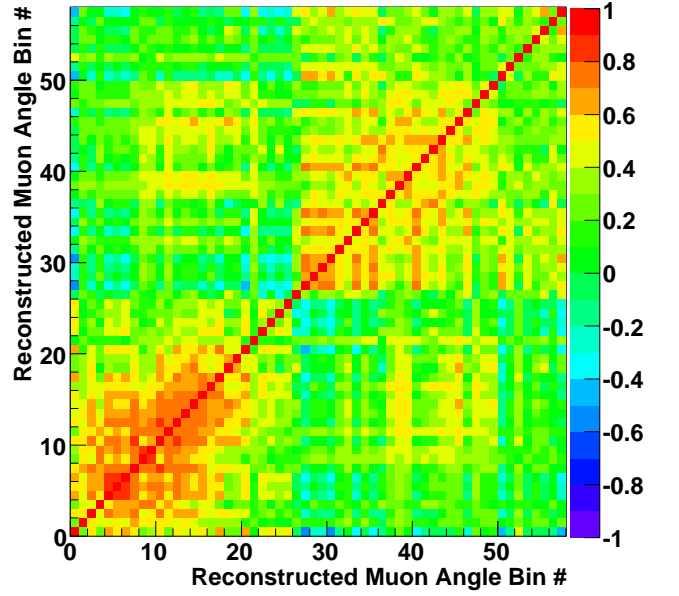}}
\subfigure[~Antineutrino Sample]{\includegraphics[width=\columnwidth]{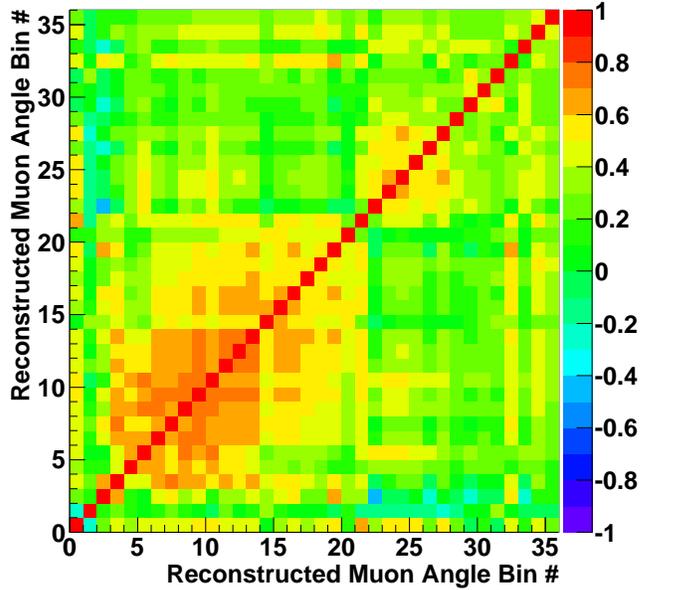}}
\caption{Correlation matrix associated with the systematic and statistical uncertainties for the neutrino and antineutrino angular distributions using NUANCE.}
\label{fig:corr_matrix_nuance}
\end{center}
\end{figure}

A summary of the fit results obtained for the $K^+$ production and rate separately for the neutrino, antineutrino and the combined neutrino and antineutrino samples is presented in Tab.~\ref{tab:nuance_fit_results} relative to the MC beam prediction. Fig.~\ref{fig:corr_matrix_nuance} reports the correlation matrices respectively for the neutrino and the antineutrino muon angular distributions. The full systematic covariance matrices for the neutrino and antineutrino angular distribution contain normalization uncertainties of 19\% and 25\%, respectively. For the initial cross section fits with pull terms, the common normalization uncertainty in the parameters is 6.5\%, but as described above, the final fit for the $K^+$ weight determination only uses the parameter values and not the uncertainties.

The average $K^+$ energy and angle for the combined neutrino and antineutrino samples are reported in Tab.~\ref{tab:neutrino+antineutrino_energy_angle_NUANCE}.

\begin{table}[htbp!]
  \begin{center}
    \caption{$K^+$ fit results for the rate and production relative to the MC beam prediction for the neutrino, antineutrino and combined neutrino and antineutrino samples including the final $\chi^2$/dof obtained from the $K^{+}$ production fit for NUANCE. Errors include statistical and systematic errors. The neutrino cross-section normalizations are held at the minimized values as listed in Table~\ref{tab:nuance_fit_results_xsect} and are relative to the NUANCE predictions.}
    \vspace{0.5cm}
    \begin{tabular}{cc|c|cc}\hline\hline
                                                &                                 		&                                		& Combined \\
                                                & $\nu$-mode             		& $\bar{\nu}$-mode   		& $\nu$+$\bar{\nu}$ mode \\ \hline
    $K^{+}$ Prod.                      & 0.89$\pm$0.04$\pm$0.12      & 0.54$\pm$0.09$\pm$0.32     & 0.85$\pm$0.04$\pm$0.11\\
    $K^{+}$ Rate                       & 0.94$\pm$0.05$\pm$0.11      & 0.54$\pm$0.09$\pm$0.30     & 0.88$\pm$0.04$\pm$0.10\\
    $\chi^2$/dof		                & 47.8/45                 			& 18.5/27 & 67.3/79            \\\hline\hline
    \end{tabular}
  \label{tab:nuance_fit_results}
  \end{center}
\end{table}


\begin{table}[htbn!]
\begin{center}
\caption{Measured $\dfrac{d^2\sigma}{dpd\Omega}$, mean energy, and mean angle (with respect to proton beam direction) for the selected $K^+$ in neutrino, antineutrino, and the combined neutrino and antineutrino samples using NUANCE. Errors on the mean energy and mean angle values correspond to the error on the mean for the relative distributions.}
\vspace{0.5cm}
\begin{tabular}{c|c|c|c}\hline\hline
                                                              & $E_{K^+}$ (GeV)   & $\theta_{K^+}$(degree) & $\dfrac{d^2\sigma}{dpd\Omega}$ \\
                                                              &                               &                                        & (mb/(GeV/c $\times$ sr) \\ \hline
$\nu$-mode                                           & 3.84$\pm$0.03          & 4.07$\pm$0.06                   &  5.77$\pm$0.83      \\
$\bar{\nu}$-mode                                  & 4.32$\pm$0.07          & 2.04$\pm$0.09                   &  3.18$\pm$1.94       \\
$\nu$ + $\bar{\nu}$-mode                     & 3.93$\pm$0.03          & 3.71$\pm$0.04                   &  5.34$\pm$0.76       \\ \hline\hline
\end{tabular}
\label{tab:neutrino+antineutrino_energy_angle_NUANCE}
\end{center}
\end{table}

The $K^+$ fit results for the production double differential cross-section in neutrino mode and antineutrino mode, though consistent, do not have to agree on the same central value. The sample of $K^+$ events selected in antineutrino mode has higher energy and lower angle with respect to the ones selected in neutrino mode as shown in Tab.~\ref{tab:numu_energy} and in Tab.~\ref{tab:antinumu_energy}. Combining the neutrino and antineutrino data gives a broader kinematical region for the $K^+$ rate and production measurements.

The values for $\dfrac{d^2\sigma}{dpd\Omega}$ for the neutrino, antineutrino, and combined mode results are given in Tab.~\ref{tab:neutrino+antineutrino_energy_angle_NUANCE} along with the mean energy and angles for the corresponding $K^+$ samples. These values are obtained multiplying the measured $K^+$ production in Tab.~\ref{tab:nuance_fit_results} by the MC beam prediction in Eq.~\ref{eq:central_value_doublediff}.

\begin{figure}[htbp!]
\begin{center}
\subfigure[~1-Track Sample]{\includegraphics[width=\columnwidth]{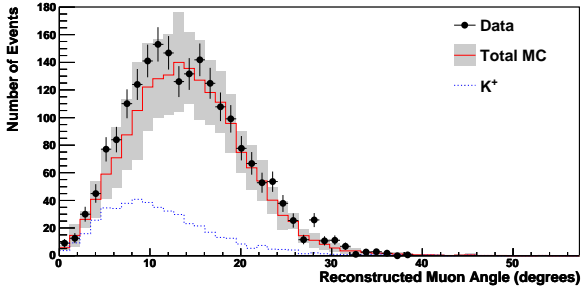}}
\subfigure[~2-Track Sample]{\includegraphics[width=\columnwidth]{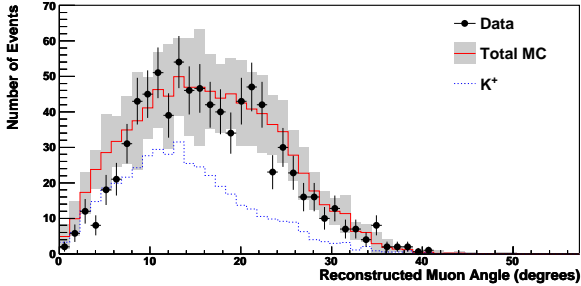}}
\subfigure[~3-Track Sample]{\includegraphics[width=\columnwidth]{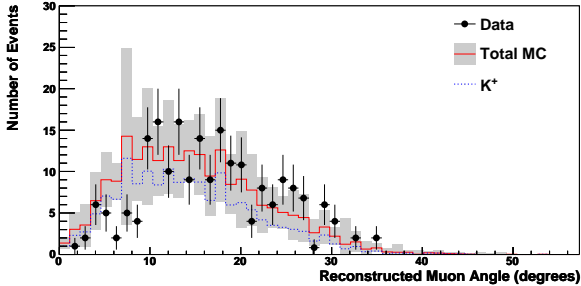}}
\caption{Reconstructed muon angle for the SciBar 1-Track, 2-Track, and 3-Track samples after the fit in neutrino mode running for NUANCE. The $K^+$ production weight and the cross-section values in Table~\ref{tab:nuance_fit_results_xsect} have been applied to the NUANCE MC predictions. The grey area represents the total systematic uncertainty in the MC.}
\label{fig:nuance_numu_selectedsample_normalized}
\end{center}
\end{figure}
\begin{figure}[htbp!]
\begin{center}
\subfigure[~1-Track Sample]{\includegraphics[width=\columnwidth]{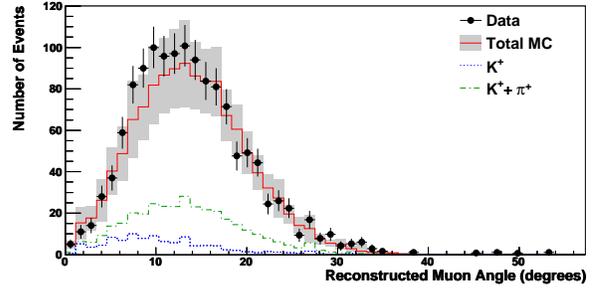}}
\subfigure[~2-Track Sample]{\includegraphics[width=\columnwidth]{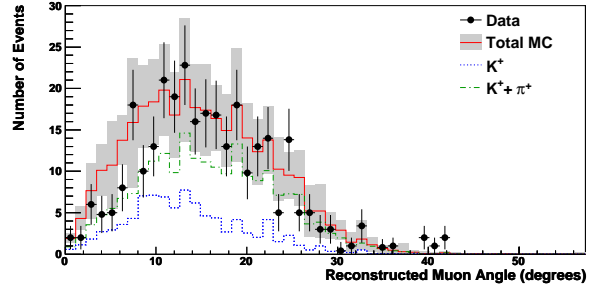}}
\subfigure[~3-Track Sample]{\includegraphics[width=\columnwidth]{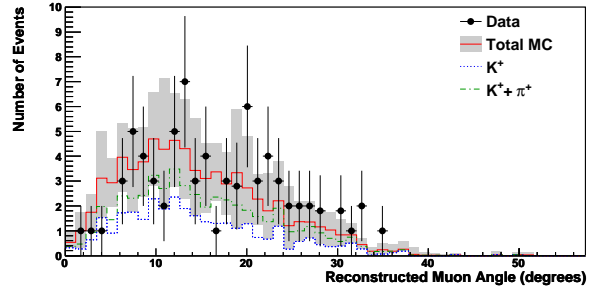}}
\caption{Reconstructed muon angle for the SciBar 1-Track, 2-Track, and 3-Track samples after the fit in antineutrino mode running for NUANCE. The $K^+$ production weight and the cross-section values in Table~\ref{tab:nuance_fit_results_xsect} have been applied to the NUANCE MC. The grey area represents the total systematic uncertainty in the MC.}
\label{fig:nuance_antinumu_selectedsample_normalized}
\end{center}
\end{figure}

The reconstructed muon angular distribution for the SciBar 1, 2 and 3-Track sample rescaled using the fit results are shown in Fig.~\ref{fig:nuance_numu_selectedsample_normalized} for neutrino mode and in Fig.~\ref{fig:nuance_antinumu_selectedsample_normalized} for antineutrino mode.

\section{Conclusion}

\par In summary, we have made a measurement of the kaon production rate and cross section for 8 GeV protons on a Be target in the Fermilab BNB using high-energy muon neutrino events observed in the SciBooNE detector. SciBooNE's full neutrino (antineutrino) data set, corresponding to $0.99 \times 10^{20}$ ($1.51\times 10^{20}$) POT, is used. The primary measurement uses the NUANCE interactions simulation and can be directly applied to the MiniBooNE oscillation analysis~\cite{AguilarArevalo:2007it} which used that simulation. A comparison of the results obtained with the NEUT neutrino interaction simulation gives similar results (see Appendix \ref{sec:neutnuance}).

\par The analysis of the neutrino and antineutrino mode data are summarized in Tab.~\ref{tab:nuance_fit_results}. Performing the analysis on the neutrino and antineutrino mode data sets yields the following measurement for $\dfrac{d^2\sigma}{dpd\Omega}$ for the $p + Be\rightarrow K^+ + X$ at the mean $K^+$ energy of 3.9~GeV and angle 3.7~degrees:
\begin{eqnarray}
\dfrac{d^2\sigma}{dpd\Omega}~= ~(5.34\pm0.76)~mb/(GeV/c \times sr).
\end{eqnarray}
The same analysis is also performed using the NEUT simulation giving the results reported in Tab.~\ref{tab:neut_fit_results} and the result is:
$\dfrac{d^2\sigma}{dpd\Omega}~= ~(5.49\pm0.79)~mb/(GeV/c \times sr)$.
As seen, the two simulations give very similar values (within 2.5\%) and demonstrate the independency of the result with respect to the underlying neutrino cross-sections and nuclear models that are different between NUANCE and NEUT.

\par Applying these measurements will significantly reduce the systematic uncertainty associated with the measured $\nu_e$ background in MiniBooNE, which is a major source of uncertainty in the previously published $\nu_e$ oscillation appearance result~\cite{AguilarArevalo:2009xn} where the MiniBooNE collaboration used a conservative 40\% error for the $K^+$ flux uncertainty.
The $K^+$ rates (production$\times$neutrino cross section) obtained using this analysis are:
\begin{eqnarray}
\nonumber K^{+}~Rate &=& 0.85 \pm 0.11,
\end{eqnarray}
for both NUANCE and NEUT.
In addition, the fact that the normalizations of the $K^+$ production and rate weights are less than one $\sigma$ away from the MC predictions (Eq.~\ref{eq:central_value_doublediff}) validates the procedure used to extrapolate the high energy kaon data to lower energy using the Feynman Scaling model \cite{shaevitz:2010}. The addition of this measurement in future fits to this model will reduce the systematic uncertainty associated with $K^+$ production in low energy neutrino beams.


\section{Acknowledgment}

We acknowledge the Physics Department at Chonnam National University,
Dongshin University, and Seoul National University for the loan of
parts used in SciBar and the help in the assembly of SciBar.
We wish to thank the Physics Departments at
the University of Rochester and Kansas State University for the loan
of Hamamatsu PMTs used in the MRD.  We gratefully acknowledge support
from Fermilab as well as various grants, contracts and fellowships
from the MEXT and JSPS (Japan), the INFN (Italy), the Ministry of Science
and Innovation and CSIC (Spain), the STFC (UK), and the DOE and NSF (USA).
This work was supported by MEXT and JSPS with the Grant-in-Aid
for Scientific Research A 19204026, Young Scientists S 20674004,
Young Scientists B 18740145, Scientific Research on Priority Areas
``New Developments of Flavor Physics'', and the global COE program
``The Next Generation of Physics, Spun from Universality and Emergence''.
The project was supported by the Japan/U.S. Cooperation Program in the field
of High Energy Physics and by JSPS and NSF under the Japan-U.S. Cooperative
Science Program.

\appendix

\section{Cross-Section Fit Details}\label{app:xsect-fit}

In this section we will describe the details of the cross-section fit used in the $\chi^2$ minimization as described in Sec.~\ref{sec:covariancefit}, Eq.~\ref{eq:chi2}.
The cross-sections described in Sec.~\ref{sec:neutrinointeractionsimulation} have been measured in the past few years with high precision by the MiniBooNE collaboration~\cite{teppei:2007ru, AguilarArevalo:2011sz, AguilarArevalo:2010bm, AguilarArevalo:2008yp}.

Eq.~\ref{eq:pred_for_production_i} and Eq.~\ref{eq:pred_for_rate_i} are used to predict the number of neutrino events in bin i in case the fit is performed to measure the $K^+$ production or the $K^+$ rate. Slightly different equations, Eq.~\ref{eq:pred_for_production_antinu} and Eq.~\ref{eq:pred_for_rate_antinu} are used in the antineutrino mode fit. $R_{K^{+}}^{Prod.}$ and $R_{K^{+}}^{Rate}$ are the fit parameters for the $K^+$ production and $K^+$ rate.

We use different normalization factors depending on the neutrino parent and on the interaction type. The normalization factors depend on the neutrino parent ($K^+$, $\pi^+$, $\pi^-$) and on the interaction types (QE, 1$\pi$, and Other). For $K^+$ production, the $R_{\nu}^{QE}$ and $R_{\nu}^{1\pi}$ is applied to both the $K^+$ and $\pi^+$ while for $K^+$ rate, the $R_{\nu}^{QE}$ and $R_{\nu}^{1\pi}$ is applied to only the $\pi^+$ (the neutrino cross-sections are implicit in the $K^+$ normalization itself). The reason we use the same normalization factors for neutrinos coming from $\pi^+$ and $K^+$ is because we assume no dependence in the cross-section at high energy ($>$~2~GeV) from the neutrino parent. We apply different normalization factors between neutrino and antineutrino mode running for neutrinos from CCQE and CC1$\pi$ interaction types to correctly take into account the MiniBooNE measurement~\cite{AguilarArevalo:2011sz} on the different neutrino flux prediction in antineutrino mode running. We use $R_{\nu}^{QE}$ and $R_{\nu}^{1\pi}$ for neutrino mode running and $R_{\nu}^{'QE}$ and $R_{\nu}^{'1\pi}$ for antineutrino mode running. We do not apply any rescaling to all the other neutrino interaction types. For neutrino mode running, we do not rescale any interactions coming from antineutrinos due to the very small contamination of antineutrinos.

\begin{eqnarray}
&&N_{i}^{pred}|_{Prod.} = \nonumber \\
&&R_{K^{+}}^{Prod.} \times (R_{\nu}^{QE} \times N_i^{K^{+}_{QE}} + R_{\nu}^{1\pi} \times N_i^{K^{+}_{1\pi}} + N_i^{K^{+}_{Other}}) \nonumber \\
&&+~R_{\nu}^{QE} \times N_i^{\pi^{+}_{QE}} + R_{\nu}^{1\pi} \times N_i^{\pi^{+}_{1\pi}} + N_i^{\pi^{+}_{Other}} \nonumber \\
&&+~N_i^{\pi^{-}_{QE}} + N_i^{\pi^{-}_{1\pi}} + N_i^{\pi^{-}_{Other}} + N_i^{Other}
\label{eq:pred_for_production_i}
\end{eqnarray}

\begin{eqnarray}
&N&_{i}^{pred}|_{Rate} = R_{K^{+}}^{Rate} \times (N_i^{K^{+}_{QE}} + N_i^{K^{+}_{1\pi}} + N_i^{K^{+}_{Other}}) \nonumber \\
&+&~R_{\nu}^{QE} \times N_i^{\pi^{+}_{QE}} + R_{\nu}^{1\pi} \times N_i^{\pi^{+}_{1\pi}} + N_i^{\pi^{+}_{Other}} \nonumber \\
&+&~N_i^{\pi^{-}_{QE}} + N_i^{\pi^{-}_{1\pi}} + N_i^{\pi^{-}_{Other}} + N_i^{Other}
\label{eq:pred_for_rate_i}
\end{eqnarray}

For antineutrino mode running, we applied additional normalization factors to take into account neutrinos produced by $\pi^-$ as shown in Eq.~\ref{eq:pred_for_production_antinu} and Eq.~\ref{eq:pred_for_rate_antinu}.

\begin{eqnarray}
&M&_{p}^{pred}|_{Prod.} = \nonumber \\
&R&_{K^{+}}^{Prod.} \times (R_{\nu}^{'QE} \times M_p^{K^{+}_{QE}} + R_{\nu}^{'1\pi} \times M_p^{K^{+}_{1\pi}} + M_p^{K^{+}_{Other}}) \nonumber \\
&+& R_{\nu}^{'QE} \times M_p^{\pi^{+}_{QE}} + R_{\nu}^{'1\pi} \times M_p^{\pi^{+}_{1\pi}} + M_p^{\pi^{+}_{Other}} \nonumber \\
&+& R_{\bar{\nu}}^{QE} \times M_p^{\pi^{-}_{QE}} + R_{\bar{\nu}}^{1\pi} \times M_p^{\pi^{-}_{1\pi}} + M_p^{\pi^{-}_{Other}} \nonumber \\
&+& M_p^{Other}
\label{eq:pred_for_production_antinu}
\end{eqnarray}

\begin{eqnarray}
&M&_{p}^{pred}|_{Rate} = R_{K^{+}}^{Rate} \times (M_p^{K^{+}_{QE}} + M_p^{K^{+}_{1\pi}} + M_p^{K^{+}_{Other}}) \nonumber \\
&+& R_{\nu}^{'QE} \times M_p^{\pi^{+}_{QE}} + R_{\nu}^{'1\pi} \times M_p^{\pi^{+}_{1\pi}} + M_p^{\pi^{+}_{Other}} \nonumber \\
&+& R_{\bar{\nu}}^{QE} \times M_p^{\pi^{-}_{QE}} + R_{\bar{\nu}}^{1\pi} \times M_p^{\pi^{-}_{1\pi}} + M_p^{\pi^{-}_{Other}} \nonumber \\
&+& M_p^{Other}
\label{eq:pred_for_rate_antinu}
\end{eqnarray}

Additional pull terms are added to the fit when we fit the $K^+$ normalization factors and the neutrino cross-sections. We add a pull term for each cross-section weight (6 terms total). These pull terms are then removed when performing the final ${\chi}^2$ minimization, and the cross section values are fixed at their best fit values. Each of the pull terms is of the form $\dfrac{(R_{CS} - \mu)^2} {\sigma^2}$, where $R_{CS}$ is the cross-section best fit value, $\mu$ is the cross-section initial value, and $\sigma$ is the cross-section uncertainty.

In the case of the antineutrino mode fit for the NUANCE MC we use a larger $R_{\bar{\nu}}^{CCQE}$ and $R_{\bar{\nu}}^{1\pi}$ with respect to the central fit values to get a better agreement between data and MC for the SciBar 1-Track sample as shown in Fig.~\ref{fig:antinumu_selectedsample}. For all other pull terms, $\mu$ is set at 1.0 with a 30$\%$ uncertainty. A summary of the cross-section best fit values is in Tab.~\ref{tab:nuance_fit_results_xsect} for NUANCE MC and in Tab.~\ref{tab:neutfitxsect} for NEUT MC. The variations in these normalization factors for the neutrino and antineutrino cross-sections are well within the cross-section systematic uncertainties and the cross-section best fit values agree well with the cross-section measurements performed by the MiniBooNE collaboration~\cite{teppei:2007ru, AguilarArevalo:2011sz, AguilarArevalo:2010bm, AguilarArevalo:2011sz}.

\section{$K^+$ Measurements using NEUT}
\label{sec:neutnuance}

In the SciBooNE experiment, neutrino interactions with carbon and hydrogen in the SciBar detector
are also simulated by the NEUT~\cite{Hayato:2002sd, Mitsuka:2008zz} program library.

\begin{figure}[htbn!]
\begin{center}
\subfigure[~1-Track Sample]{\includegraphics[width=\columnwidth]{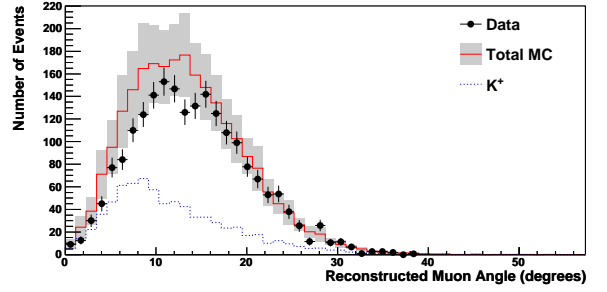}}
\subfigure[~2-Track Sample]{\includegraphics[width=\columnwidth]{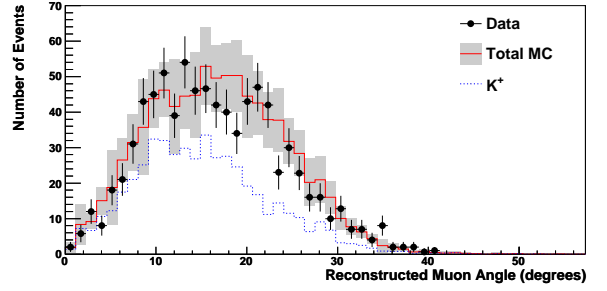}}
\subfigure[~3-Track Sample]{\includegraphics[width=\columnwidth]{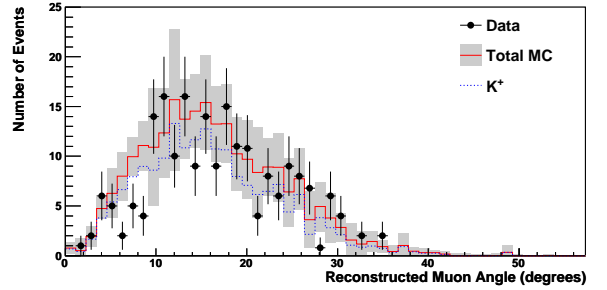}}
\caption{Reconstructed muon angle for the SciBar 1-Track, 2-Track, and 3-Track samples in neutrino mode running for data and NEUT MC. The background contributions from $K^{-}$ and $K^{0}_{L}$ are very small but included in the MC histogram. The grey area represents the total systematic uncertainty in the MC.}
\label{fig:neut_selected_sample}
\end{center}
\end{figure}

\begin{figure}[htbn!]
\begin{center}
\subfigure[~1-Track Sample]{\includegraphics[width=\columnwidth]{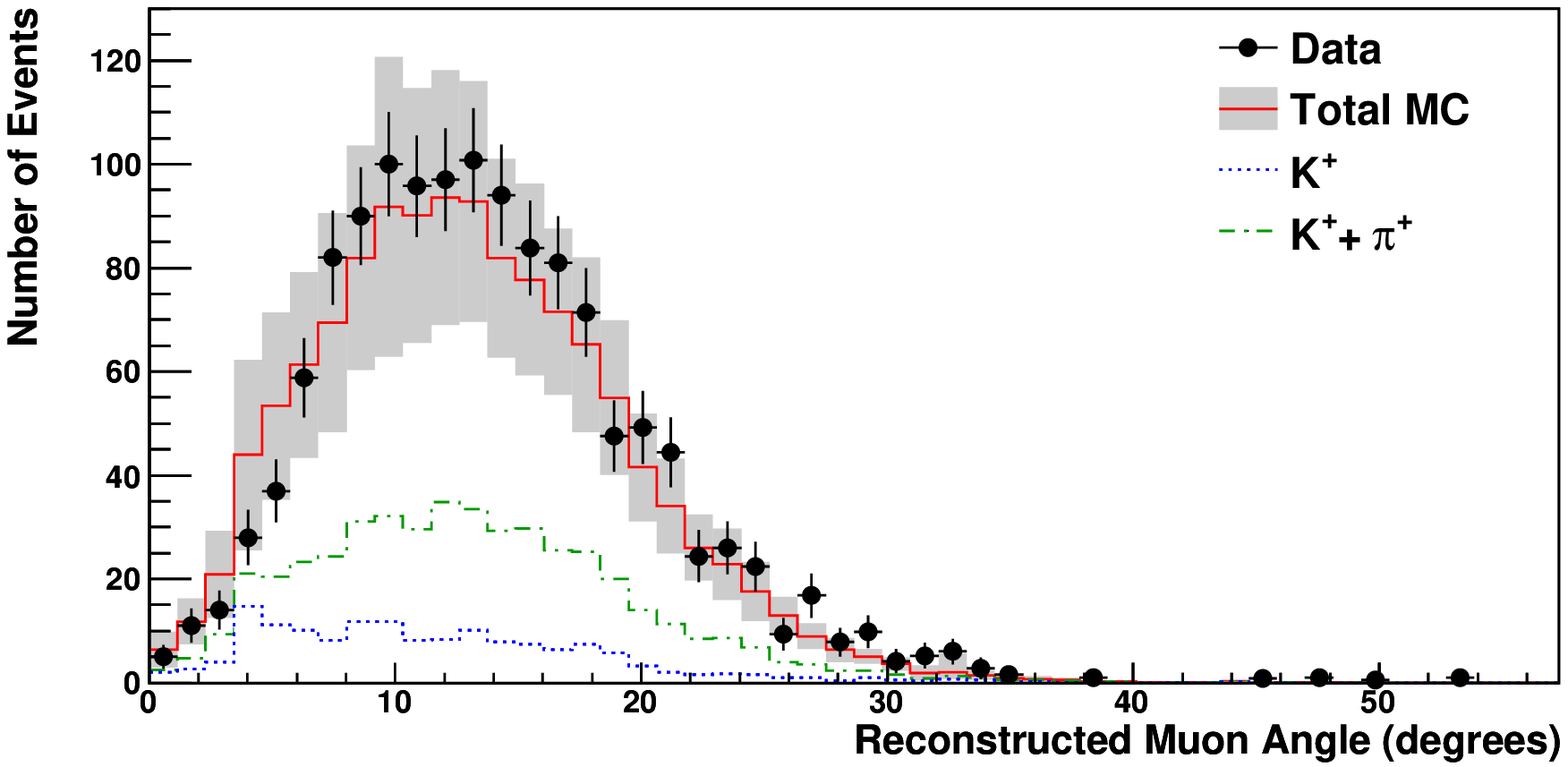}}
\subfigure[~2-Track Sample]{\includegraphics[width=\columnwidth]{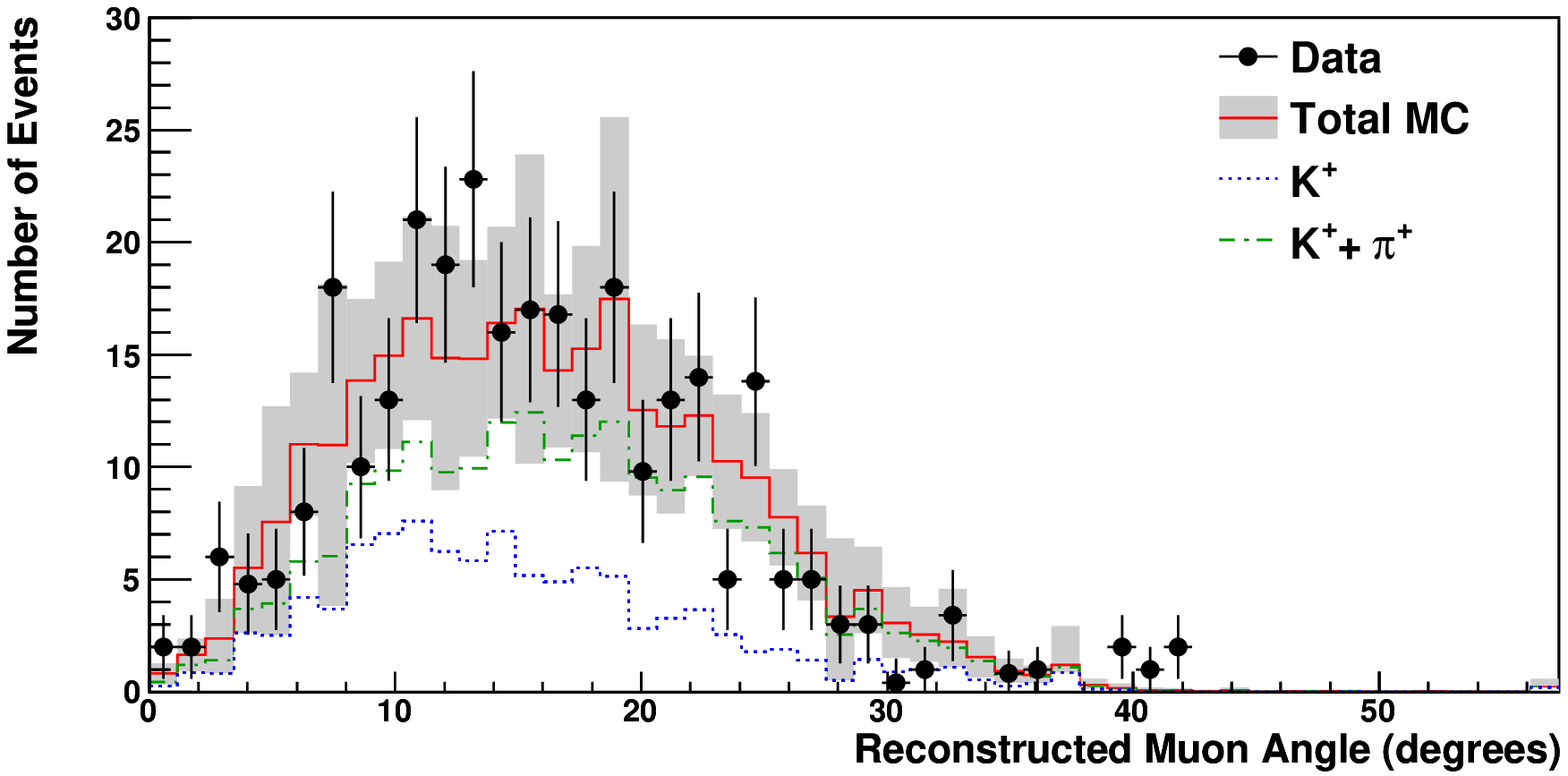}}
\subfigure[~3-Track Sample]{\includegraphics[width=\columnwidth]{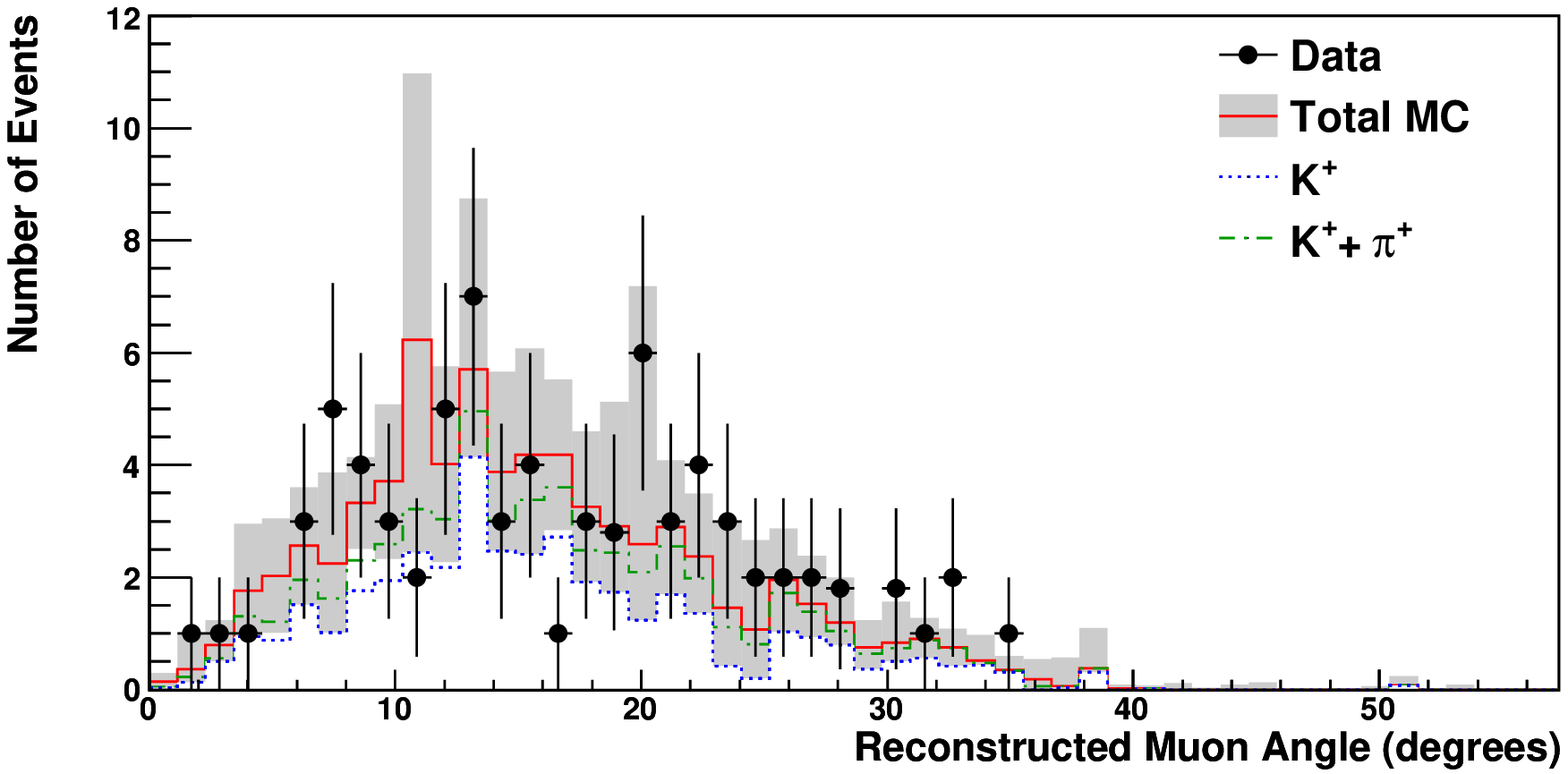}}
\caption{Reconstructed muon angle for the SciBar 1-Track, 2-Track, and 3-Track samples in antineutrino mode running for data and NEUT MC. The background contributions from $K^{-}$ and $K^{0}_{L}$ are small but included in the MC histogram. The grey area represents the total systematic uncertainty in the MC.}
\label{fig:neut_antinumu_selectedsample}
\end{center}
\end{figure}

The measurements presented in Sec.~\ref{sec:muon_neutrino_analysis} and Sec.~\ref{sec:antinumu_analyisis} are computed with assumptions on the underlying neutrino cross sections and nuclear models. Due to that, we have repeated our analysis using NEUT~\cite{Hayato:2002sd,Mitsuka:2008zz} as neutrino interactions Monte Carlo. The analysis is performed in the same way as for the NUANCE MC. The reconstructed muon angle distributions for the SciBar 1, 2 and 3-Track samples are shown in Fig.~\ref{fig:neut_selected_sample} for the neutrino mode analysis and in Fig.~\ref{fig:neut_antinumu_selectedsample} for the antineutrino mode analysis.

NEUT and NUANCE parameters are summarized in Tab.~\ref{tab:xsec_params_neutnuance}. The different parameter values result in different neutrino cross section predictions between the two and the differences are more in how the two Monte Carlos implement nuclear and cross section models. The total interaction numbers predicted by NEUT are summarized in Tab.~\ref{tab:neut_interactions}.

\begin{table}[htbp!]
  \begin{center}
  \caption{Parameter used for Neutrino interaction simulation.}
  \vspace{0.5cm}
  \begin{tabular}{cccc}\hline\hline
    Parameter  & NEUT & NUANCE  \\
    \hline
    $p_{F}$   & 217 MeV &  220 MeV\\
    $E_{B}$   & 25 MeV  &  34 MeV \\
    $M_A^{QE}$-C     & 1.21 GeV  &  1.234 GeV\\
    $M_A^{QE}$-H     & 1.0 GeV  &  1.0 GeV\\
    $\kappa$  & 1.00      &  1.022    \\
    $M_A^{1\pi}$ & 1.21  GeV &1.10 GeV\\
    $M_A^{coherent}$ & 1.03  GeV &1.03 GeV\\
    $M_A^{N\pi}$ & (DIS) &1.3 GeV\\\hline
  \end{tabular}
  \label{tab:xsec_params_neutnuance}
  \end{center}
\end{table}

\begin{table}[htbn!]
\begin{center}
\caption{The expected number of events in each neutrino interaction estimated by NEUT at the SciBooNE detector location with the neutrino beam exposure of $0.99 \times 10^{20}$ protons on target for neutrino mode and of $1.51 \times 10^{20}$ for antineutrino mode. The 9.8~ton fiducial volume of the SciBar detector is assumed. Charged current and neutral current interactions are abbreviated as CC and NC, respectively.}
\vspace{0.5cm}
 \begin{tabular}{lcc}\hline\hline
                                    &  \multicolumn{2}{c}{NEUT} \\
                                       &  \# Events ~~~~& \# Events \\
   Interaction                   & neutrino mode ~~~~& antineutrino mode\\
   \hline
    CC QE                        &  50,330 ~~~~ &   18,804         \\
    CC single-$\pi$           &  27,836 ~~~~ &   7,350          \\
    CC coh. $\pi$              &   1,692 ~~~~ &   1,308          \\
    CC DIS+Other             &   6,445 ~~~~ &   1,760          \\
    NC                              &  35,153 ~~~~ &   13,194          \\\hline\hline
 \end{tabular}
\label{tab:neut_interactions}
\end{center}
\end{table}

The difference in the parameter choices (Tab.~\ref{tab:xsec_params_neutnuance}), leads to NEUT predicting a larger QE and single pion rate than NUANCE. This difference in QE rate comes largely from the choice of $\kappa$ values, and the difference in single pion rate can be largely accounted for by the difference in $M_A^{1\pi}$ assumptions. CC-QE interactions, which are implemented using the Smith and Moniz model~\cite{Smith:1972xh}, are the dominant interaction in the SciBooNE neutrino energy range.  The nucleons are treated as quasi-free particles and the Fermi motion of nucleons along with the Pauli exclusion principle is taken into account.  The Fermi surface momentum ($p_F$) for carbon is set to 217(220)~MeV/$c$ and the nuclear potential ($E_B$) is set to 25(34)~MeV/$c$ for NEUT(NUANCE), as extracted from electron scattering data~\cite{Moniz:1971mt}. The default binding energy in NUANCE is somewhat higher than in NEUT because it additionally accounts for neutrino vs. electron scattering differences~\cite{AguilarArevalo:2008fb}. With regards to the vector form factor, NEUT uses a dipole form with a vector mass of 0.84~$\rm GeV/c^2$, while NUANCE uses the BBA-2003 form factor~\cite{Budd:2003wb}. A dipole form is used for  the axial form factor with an adjustable axial mass, $M_A^{QE}$, for both NEUT and NUANCE.  The values of $M_A^{QE} =1.21$~$\rm GeV/c^2$ and $\kappa = 1.000$ (i.e. no additional Pauli blocking adjustment) are used in NEUT, and $M_A^{QE} = 1.234$ $\rm GeV/c^2$ and $\kappa = 1.022$ are used in NUANCE~\cite{AguilarArevalo:2008fb}. The same Fermi momentum distribution and nuclear potential are used in all other neutrino-nucleus interactions except for resonant $\pi$ production.

The resonant production of single pion, kaon, and eta mesons, as described by the model of Rein and Sehgal (RS)~\cite{Rein:1980wg}, is implemented with axial-vector form factors formalized to be dipole with $M_A^{1\pi}=1.21$~$\rm GeV/c^2$ for NEUT and $M_A^{1\pi}=1.10$~$\rm GeV/c^2$ for NUANCE. Resonance decays leading to multi-pion final states are also included in the NUANCE model as described in Sec.~\ref{sec:neutrinointeractionsimulation}. In NEUT, multi-pion production is simulated as deep inelastic scattering (DIS) and the RS model is not used.  Multi-hadron final states are simulated by two models: a custom-made program~\cite{Nakahata:1986zp} for events with W between 1.3 and 2.0~GeV/$c^2$ and PYTHIA/JETSET~\cite{Sjostrand:1993yb} for events with W larger than 2~GeV/$c^2$. The pion multiplicity is additionally restricted to be greater than one for $1.3<W<2$~GeV/$c^2$ to avoid double-counting sources of single pion production. In NUANCE, the DIS contribution slowly increases for $W$ values starting at 1.7~GeV and becomes the only source of neutrino interactions above $W>2$~GeV.

The total and inelastic pion-nucleon cross sections are treated differently in NEUT and NUANCE. In NEUT they are from the original Rein and Sehgal publication ~\cite{Rein:1982pf,Rein:2006di} while in NUANCE, they are obtained from fits to PDG data~\cite{Nakamura:2010zzi} and implemented as a function of pion energy with an additional rescale of the NC and CC coherent pion production cross section.

Intra nuclear interactions are simulated differently as well: more details can be found for both NEUT and NUANCE in~\cite{Casper:2002sd,Hiraide:2008eu}.

The results obtained for the cross-section best fit values using NEUT in the neutrino and antineutrino mode are summarized in Tab.~\ref{tab:neutfitxsect}.

\begin{table}[htbp!]
  \begin{center}
  \caption{Summary of the fit results for the cross-section normalization factors as described in Appendix \ref{app:xsect-fit} with respect to NEUT predictions. The top four values are cross section normalization values for $\nu_\mu$ and $\bar{\nu}_\mu$ coming from $\pi^+$ and $K^+$ while the last two are normalization factors for $\bar{\nu}_{\mu}$ coming from $\pi^-$.}
  \vspace{0.5cm}
  \begin{tabular}{c|c|c}\hline\hline
    $\pi^+$/$K^+$ & & Fit Value \\\hline
    & CCQE in $\nu$ mode   & 0.92$\pm$0.13 \\
    & CCQE in ${\bar{\nu}}$ mode  & 1.05$\pm$0.25 \\
    & CC1$\pi$ in $\nu$ mode  & 1.12$\pm$0.25 \\
    & CC1$\pi$ in ${\bar{\nu}}$ mode  & 0.94$\pm$0.27 \\\hline
    $\pi^-$ && \\\hline
    & CCQE in ${\bar{\nu}}$ mode & 1.01$\pm$0.30 \\
    & CC1$\pi$ in ${\bar{\nu}}$ mode & 1.01$\pm$0.30 \\
    \hline\hline
  \end{tabular}
  \label{tab:neutfitxsect}
  \end{center}
\end{table}

\begin{figure}[htbp!]
\begin{center}
\subfigure[~Neutrino Sample]{\includegraphics[width=\columnwidth]{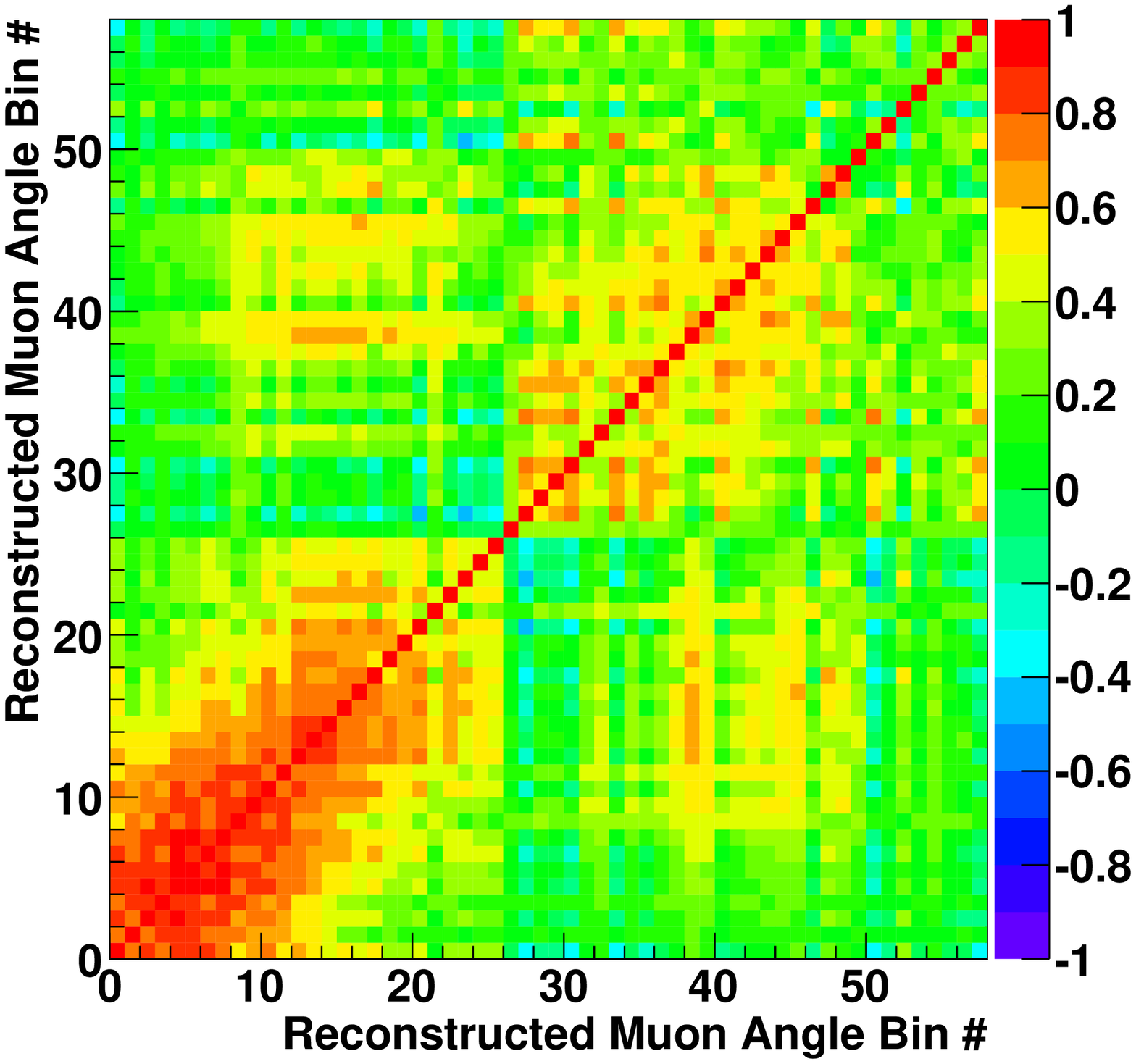}}
\subfigure[~Antineutrino Sample]{\includegraphics[width=\columnwidth]{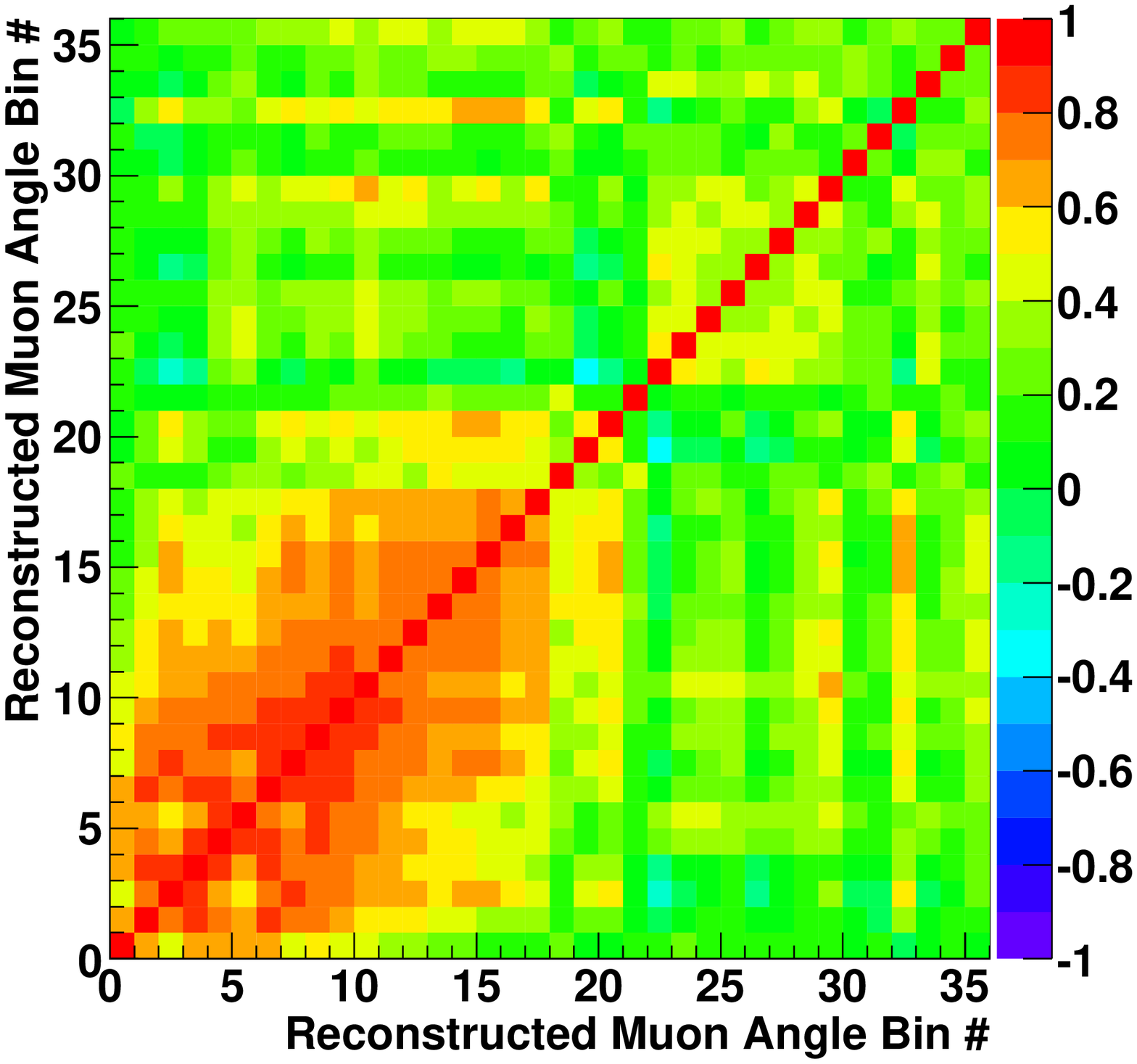}}
\caption{Correlation matrix associated with the systematic and statistical uncertainties for the neutrino and antineutrino angular distributions for NEUT.}
\label{fig:corr_matrix_neut}
\end{center}
\end{figure}

The mean energy and mean angle (with respect to proton beam direction) for the selected $K^+$ and the mean energy for $\nu_\mu$ from the selected $K^+$ in each of the three samples, for both neutrino mode running and antineutrino mode running, are summarized in Tab.~\ref{tab:k_energy_angle_summary_neut} and Tab.~\ref{tab:anti_k_energy_angle_summary_neut}.
A summary of the fit results obtained for the $K^+$ production and rate separately for neutrino, antineutrino and the combined neutrino and antineutrino samples is presented in Tab.~\ref{tab:neut_fit_results} relative to the MC beam prediction. Fig.~\ref{fig:corr_matrix_neut} reports the correlation matrices respectively for the neutrino and the antineutrino muon angular distributions. For NEUT, the full systematic covariance matrices for the neutrino and antineutrino angular distribution contain normalization uncertainties of 19\% and 25\%, respectively. For the initial cross section fits with pull terms, the common normalization uncertainty in the parameters is 6.5\%.

\begin{table}[htbn!]
\begin{center}
\caption{NEUT MC predicted mean energy and mean angle (with respect to the proton beam direction) for the selected $K^+$ samples and predicted MC energy for the selected $\nu_\mu$ from $K^+$ in the three SciBar samples in neutrino mode running. Errors correspond to the error on the mean values. The RMS of the relative distributions is reported in square parenthesis.}
\vspace{0.5cm}
\begin{tabular}{c|c|c|c}\hline\hline
         & $E_{K^+}$[RMS](GeV) & $\theta_{K^+}$[RMS](deg)  & $E_{\nu_\mu}$[RMS](GeV) \\ \hline
1-Trk  & 3.6$\pm$0.1[1.2]     & 4.2$\pm$0.1[2.0]             & 3.1$\pm$0.1[1.0]   \\ \hline
2-Trk  & 3.9$\pm$0.1[1.2]     & 4.0$\pm$0.1[2.0]             & 3.3$\pm$0.1[1.0]   \\ \hline
3-Trk  & 4.2$\pm$0.1[1.1]     & 3.6$\pm$0.1[1.9]             & 3.6$\pm$0.1[0.9]   \\ \hline\hline
\end{tabular}
\label{tab:k_energy_angle_summary_neut}
\end{center}
\end{table}

\begin{table}[htbn!]
\begin{center}
\caption{NEUT MC predicted mean energy and mean angle (with respect to the proton beam direction) for the selected $K^+$ samples and predicted MC energy for the selected $\nu_\mu$ from $K^+$ for the three SciBar samples in antineutrino mode running. Errors correspond to the error on the mean values. The RMS of the relative distributions is reported in square parenthesis.}
\vspace{0.5cm}
\begin{tabular}{c|c|c|c}\hline\hline
         & $E_{K^+}$[RMS](GeV) & $\theta_{K^+}$[RMS](deg)  & $E_{\nu_\mu}$[RMS](GeV) \\ \hline
1-Trk  & 4.1$\pm$0.1[1.3]     & 2.2$\pm$0.2[1.6]             & 3.1$\pm$0.1[1.1]   \\ \hline
2-Trk  & 4.3$\pm$0.1[1.2]     & 1.7$\pm$0.1[1.2]             & 3.4$\pm$0.1[1.0]   \\ \hline
3-Trk  & 4.7$\pm$0.2[1.1]     & 1.6$\pm$0.2[1.0]             & 3.8$\pm$0.2[1.0]   \\ \hline\hline
\end{tabular}
\label{tab:anti_k_energy_angle_summary_neut}
\end{center}
\end{table}
The average $K^+$ energy and angle for the combined neutrino and antineutrino samples are reported in Tab.~\ref{tab:neutrino+antineutrino_energy_angle_NEUT}.
\begin{table}[htbp!]
  \begin{center}
   \caption{$K^+$ fit results for the rate and production relative to the MC beam prediction for the neutrino, antineutrino and combined neutrino and antineutrino samples including the final $\chi^2$/dof obtained from the $K^{+}$ production fit for NEUT. Errors are statistical and systematic errors. The neutrino cross-section normalizations are held at the minimized values as listed in Table~\ref{tab:neutfitxsect} and are relative to the NEUT predictions.}
   \vspace{0.5cm}
     \begin{tabular}{cc|c|cc}\hline\hline
                                                         &                                 		&                                		& Combined \\
                                                         & $\nu$-mode            		& $\bar{\nu}$-mode   		& $\nu$+$\bar{\nu}$ mode \\\hline
    $K^{+}$ Prod.                               & 0.90$\pm$0.05$\pm$0.13     & 0.77$\pm$0.12$\pm$0.31     & 0.87$\pm$0.05$\pm$0.11  \\
    $K^{+}$ Rate                                & 0.92$\pm$0.05$\pm$0.11     & 0.76$\pm$0.12$\pm$0.27     & 0.88$\pm$0.05$\pm$0.10 \\
    $\chi^2$/dof			                & 40.6/45                   		& 17.7/26                  		& 58.9/77              \\\hline\hline
    \end{tabular}
    \label{tab:neut_fit_results}
  \end{center}
\end{table}
\begin{table}[htbn!]
\begin{center}
\caption{Measured $\dfrac{d^2\sigma}{dpd\Omega}$, mean energy, and mean angle (with respect to proton beam direction) for the selected $K^+$ in neutrino, antineutrino, and the combined neutrino and antineutrino samples using NEUT. Errors on the mean energy and mean angle values correspond to the error on the mean values of the relative distributions.}
\vspace{0.5cm}
\begin{tabular}{c|c|c|c}\hline\hline
                                                              & $E_{K^+}$ (GeV)   & $\theta_{K^+}$(degree) & $\dfrac{d^2\sigma}{dpd\Omega}$ \\
                                                              &                               &                                        & (mb/(GeV/c $\times$ sr) \\ \hline
$\nu$-mode                                           & 3.85$\pm$0.03          & 4.02$\pm$0.06                   &   5.85$\pm$0.90    \\
$\bar{\nu}$-mode                                  & 4.29$\pm$0.07          & 1.93$\pm$0.09                   &   4.59$\pm$1.97    \\
$\nu$ + $\bar{\nu}$-mode                     & 3.92$\pm$0.03          & 3.66$\pm$0.04                   &   5.49$\pm$0.79    \\ \hline\hline
\end{tabular}
\label{tab:neutrino+antineutrino_energy_angle_NEUT}
\end{center}
\end{table}

The values for $\dfrac{d^2\sigma}{dpd\Omega}$ for the neutrino, antineutrino, and combined mode results are given in Tab.~\ref{tab:neutrino+antineutrino_energy_angle_NEUT} along with the mean energy and angles for the corresponding $K^+$ samples.  These values are obtained multiplying the measured $K^+$ production in Tab.~\ref{tab:neut_fit_results} by the MC beam prediction in Eq.~\ref{eq:central_value_doublediff}.

\begin{figure}[htbp!]
\begin{center}
\subfigure[~1-Track Sample]{\includegraphics[width=\columnwidth]{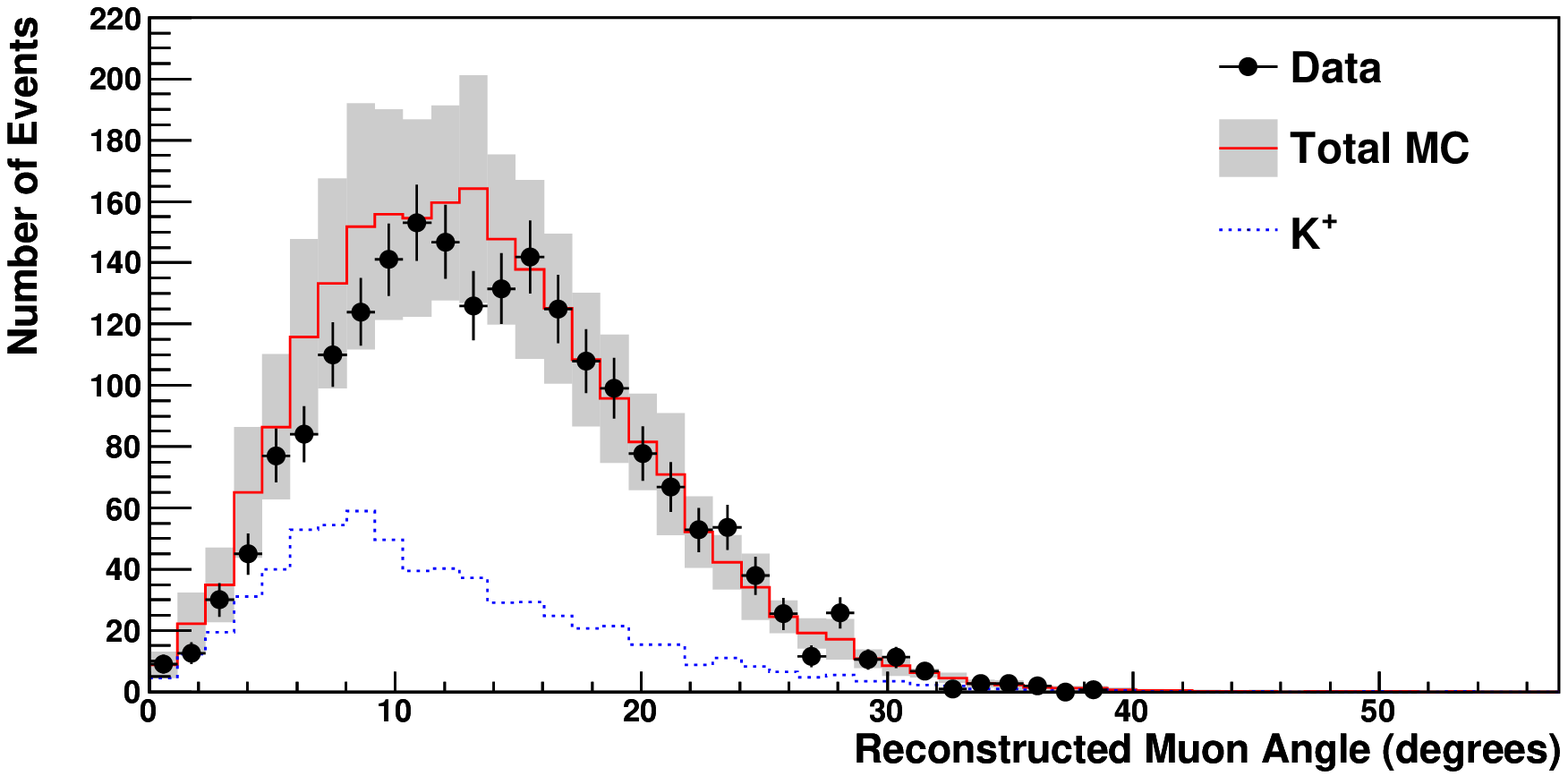}}
\subfigure[~2-Track Sample]{\includegraphics[width=\columnwidth]{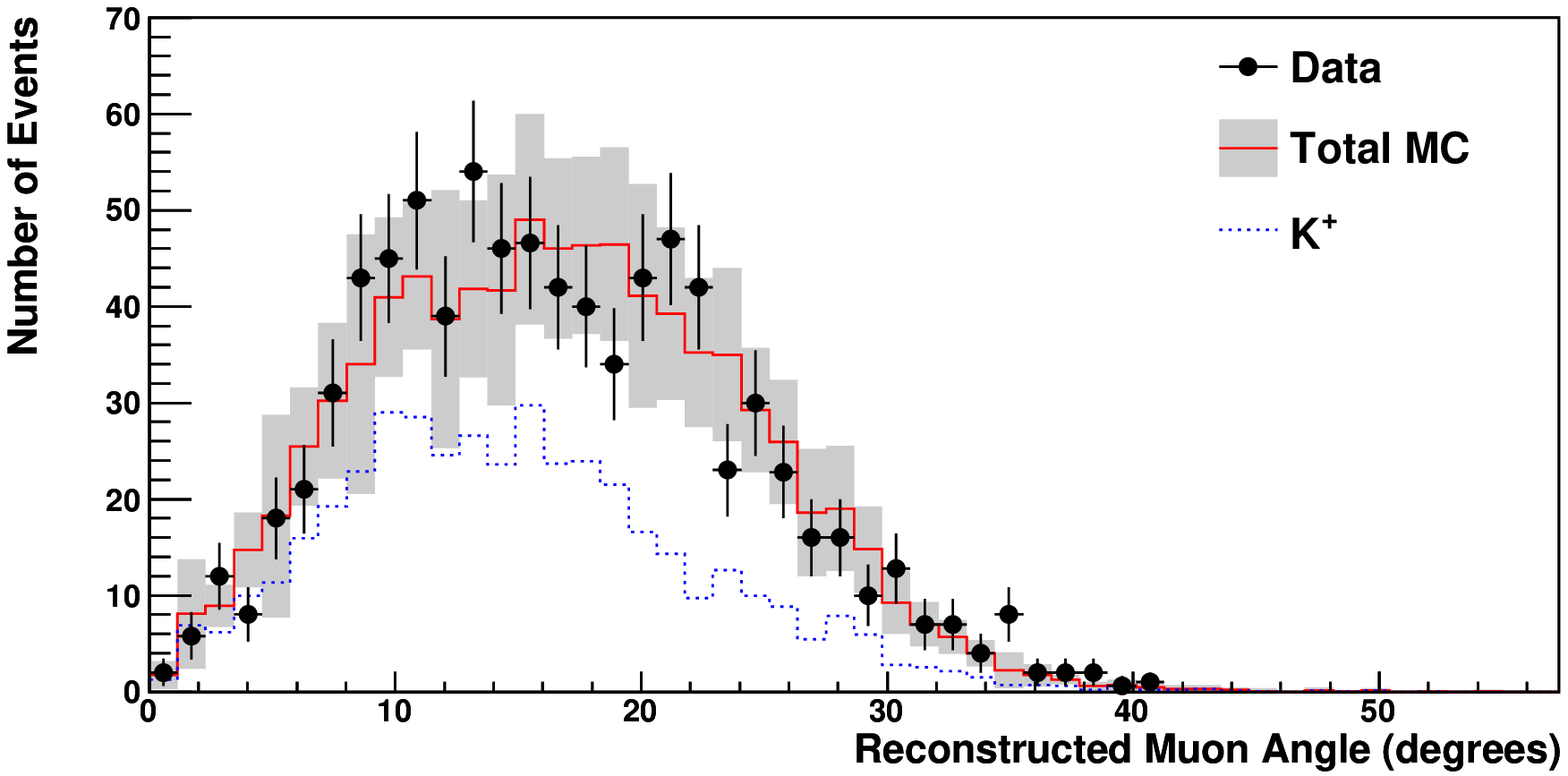}}
\subfigure[~3-Track Sample]{\includegraphics[width=\columnwidth]{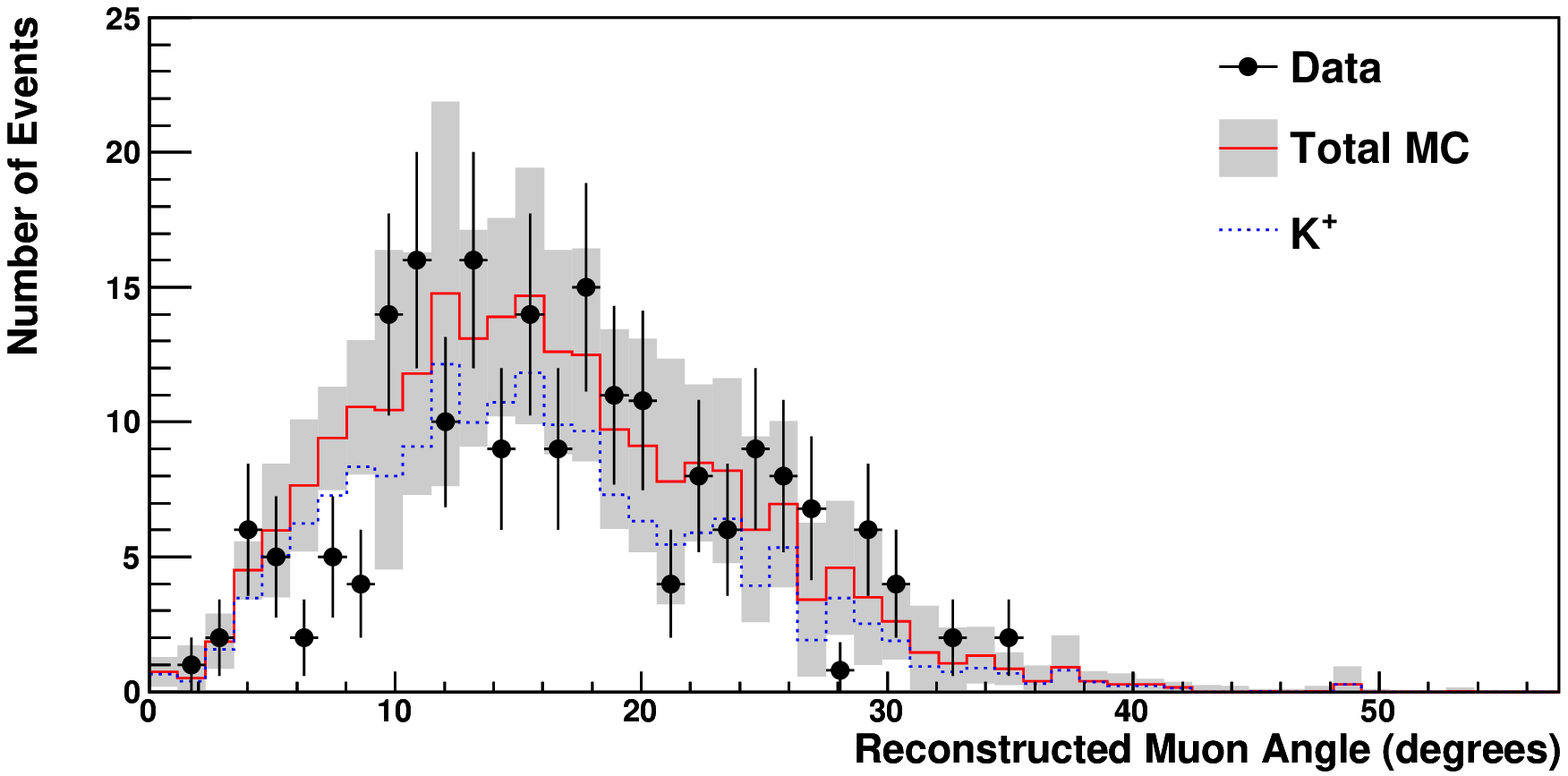}}
\caption{Reconstructed muon angle for the SciBar 1-Track, 2-Track, and 3-Track samples in neutrino mode running for data and the NEUT MC. The $K^+$ production weight and the cross-section central values in Table~\ref{tab:neutfitxsect} have been applied to the NEUT predictions. The grey area represents the total systematic uncertainty in the MC.}
\label{fig:neut_numu_selectedsample_normalized}
\end{center}
\end{figure}

\begin{figure}[htbp!]
\begin{center}
\subfigure[~1-Track Sample]{\includegraphics[width=\columnwidth]{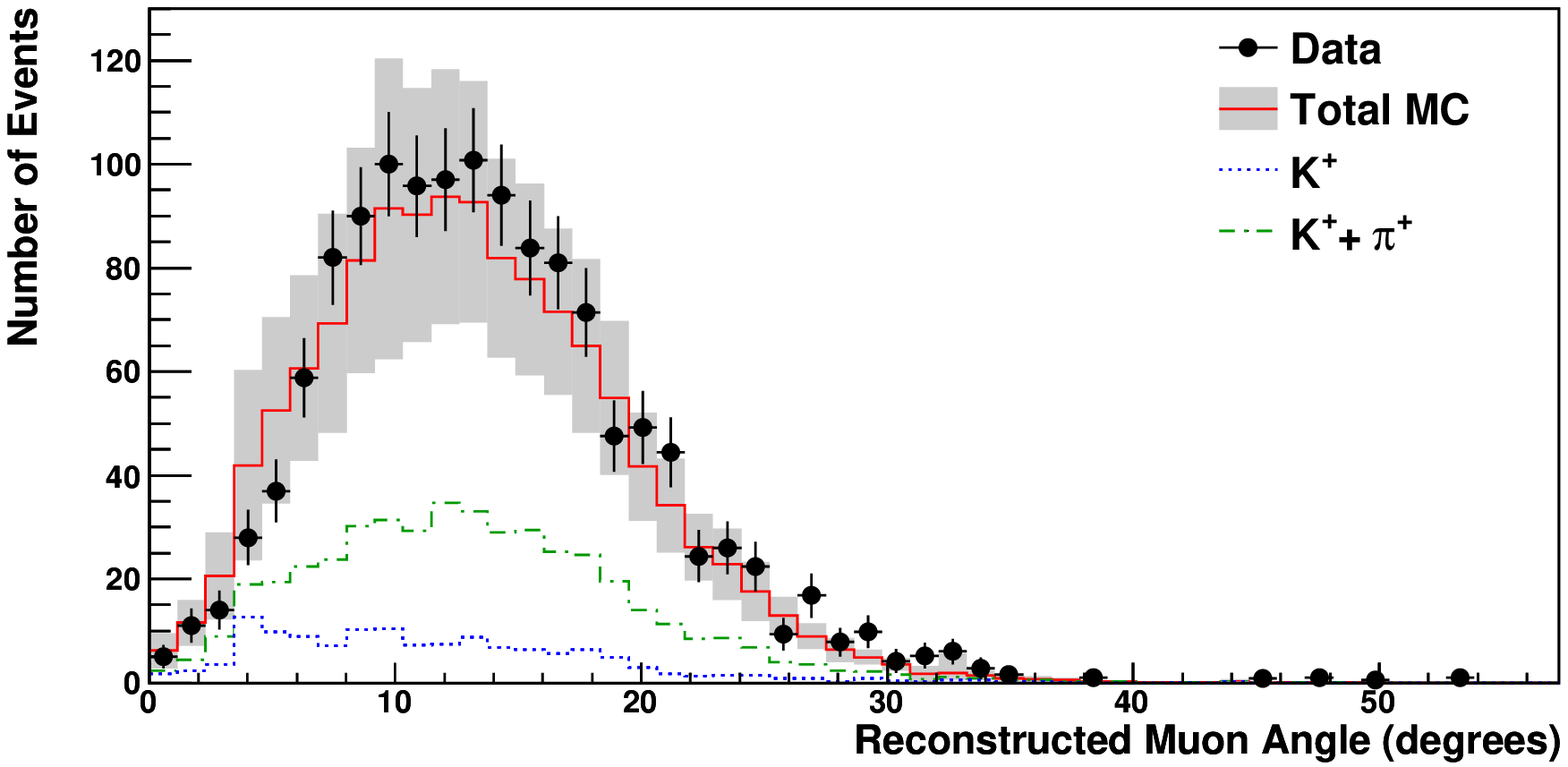}}
\subfigure[~2-Track Sample]{\includegraphics[width=\columnwidth]{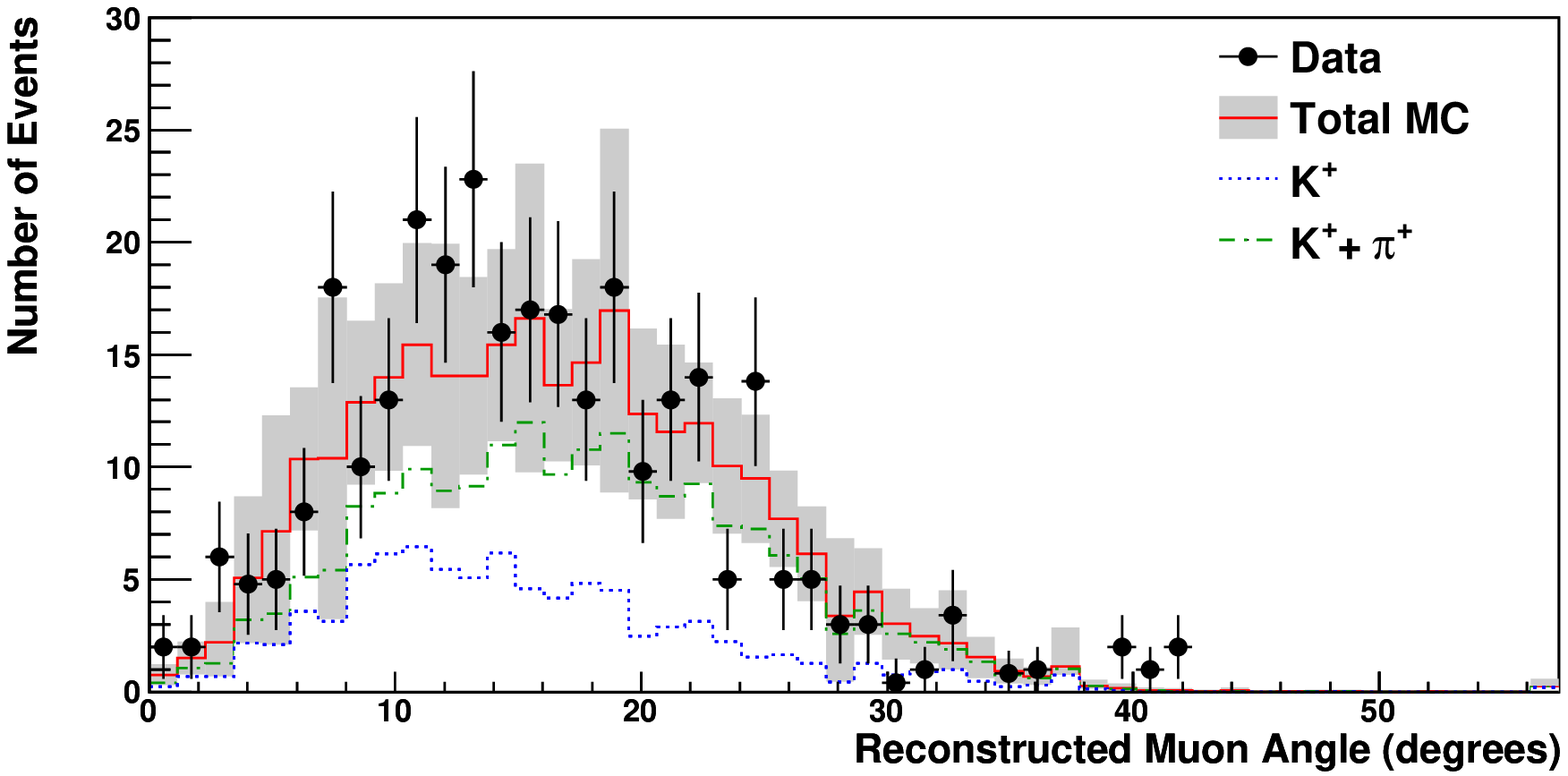}}
\subfigure[~3-Track Sample]{\includegraphics[width=\columnwidth]{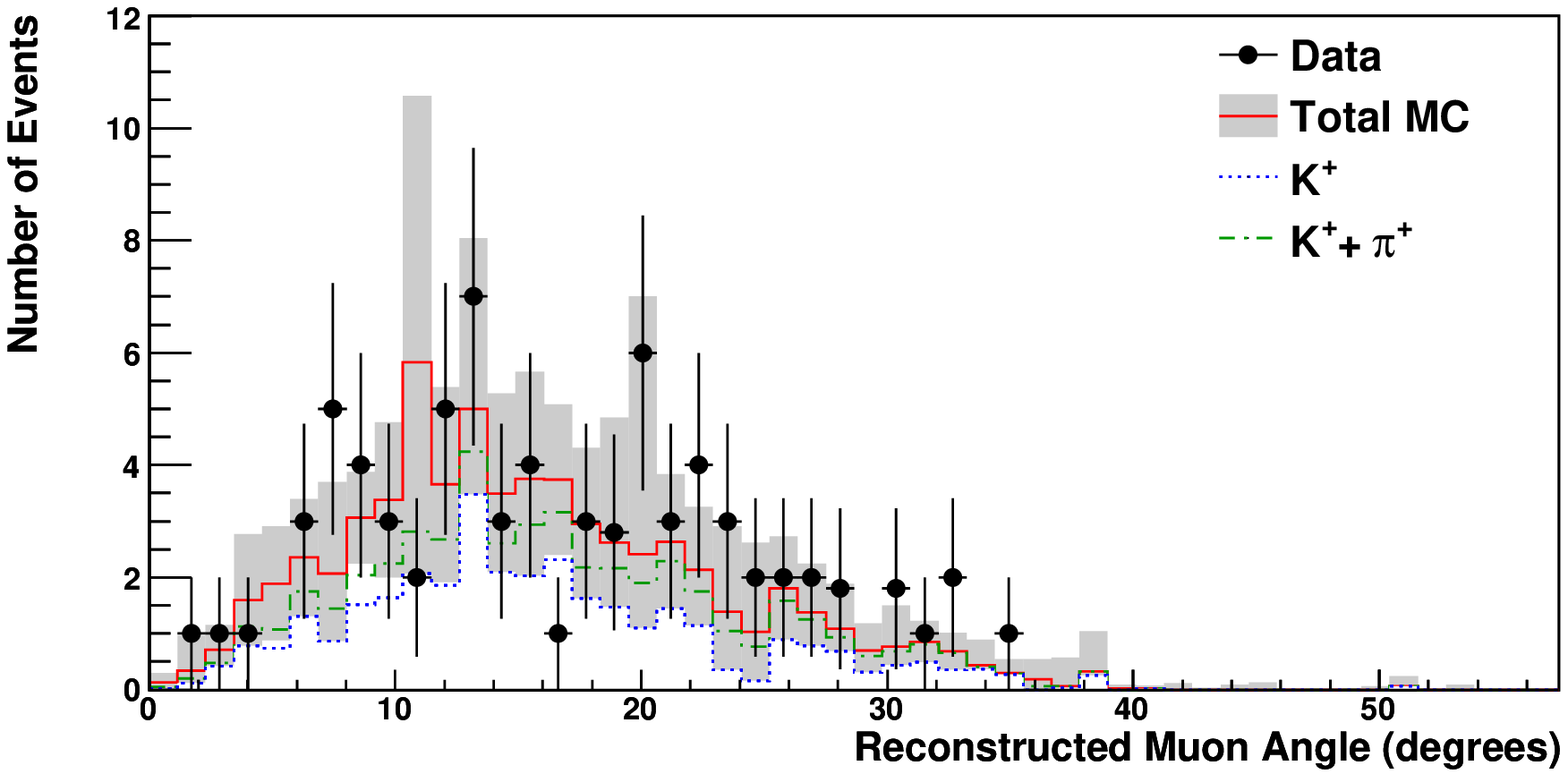}}
\caption{Reconstructed muon angle for the SciBar 1-Track, 2-Track, and 3-Track samples in antineutrino mode running for data and NEUT MC. The $K^+$ production weight and the cross-section central values in Table~\ref{tab:neutfitxsect} have been applied to the NEUT predictions. The grey area represents the total systematic uncertainty in the MC.}
\label{fig:neut_antinumu_selectedsample_normalized}
\end{center}
\end{figure}

The reconstructed muon angle distributions for the SciBar 1, 2 and 3-Track samples rescaled using the fit results are shown in Fig.~\ref{fig:neut_numu_selectedsample_normalized} for the neutrino mode analysis and in Fig.~\ref{fig:neut_antinumu_selectedsample_normalized} for the antineutrino mode analysis.

\clearpage

\bibliographystyle{unsrt}

\bibliography{reference}

\begin{thebibliography}{10}

\bibitem{Feyman::1969}
R.~P. Feynman.
\newblock Very high-energy collisions of hadrons.
\newblock {\em Phys. Rev. Lett.}, 23(24):1415--1417, Dec 1969.

\bibitem{AguilarArevalo:2008yp}
A.~A. Aguilar-Arevalo et~al.
\newblock {The Neutrino Flux prediction at MiniBooNE}.
\newblock {\em Phys. Rev.}, D79:072002, 2009.

\bibitem{Hiraide:2006zq}
K.~Hiraide.
\newblock {The SciBar detector at FNAL booster neutrino experiment}.
\newblock {\em Nucl. Phys. Proc. Suppl.}, 159:85--90, 2006.

\bibitem{Soderberg:2009rz}
Mitchell Soderberg.
\newblock {MicroBooNE: A New Liquid Argon Time Projection Chamber Experiment}.
\newblock {\em AIP Conf. Proc.}, 1189:83--87, 2009.

\bibitem{AguilarArevalo:2009xn}
A.~A. Aguilar-Arevalo et~al.
\newblock {A Search for Electron Antineutrino Appearance at the $\Delta m^2
  \sim$ 1 $\mathrm{eV}^{2}$ Scale}.
\newblock {\em Phys. Rev. Lett.}, 103:111801, 2009.

\bibitem{shaevitz:2010}
C.~Mariani et~al.
\newblock Parameterizing the p-be k$^{+}$ production cross section using a
  feynman scaling function, 2010.
\newblock PRD in preparation.

\bibitem{Sanford:1967}
J.~R. Sanford and C.~L. Wang, 1967.
\newblock BNL Internal Note No. 11299.

\bibitem{:2007gt}
M.~G. Catanesi et~al.
\newblock {Measurement of the production cross-section of positive pions in the
  collision of 8.9 GeV/c protons on beryllium}.
\newblock {\em Eur. Phys. J.}, C52:29--53, 2007.

\bibitem{Mokhov:1998kc}
N.~V. Mokhov et~al.
\newblock {MARS Code Developments}.
\newblock {\em arXiv:nucl-th/9812038}, 1998.

\bibitem{Nakajima:2011zza}
Yasuhiro Nakajima.
\newblock {A Measurement of Neutrino Charged Current Interactions and a Search
  for Muon Neutrino Disappearance with the Fermilab Booster Neutrino Beam}.
\newblock FERMILAB-THESIS-2011-09.

\bibitem{Nakajima:2010fp}
Y.~Nakajima et~al.
\newblock {Measurement of inclusive charged current interactions on carbon in a
  few-GeV neutrino beam}.
\newblock {\em Phys. Rev.}, D83:012005, 2011.

\bibitem{Nitta:2004nt}
K.~Nitta et~al.
\newblock {The K2K SciBar detector}.
\newblock {\em Nucl. Instrum. Meth.}, A535:147--151, 2004.

\bibitem{Yoshida:2004mh}
M.~Yoshida et~al.
\newblock {Development of the readout system for the K2K SciBar detector}.
\newblock {\em IEEE Trans. Nucl. Sci.}, 51:3043--3046, 2004.

\bibitem{Mariani:2009zza}
C.~Mariani.
\newblock {EC detector at SciBooNE}.
\newblock {\em J. Phys. Conf. Ser.}, 160:012035, 2009.

\bibitem{Heikkinen:2003sc}
A.~Heikkinen, N.~Stepanov, and J.~P. Wellisch.
\newblock {Bertini intra-nuclear cascade implementation in Geant4}.
\newblock {\em arXiv:nucl-th/0306008}, 2003.

\bibitem{Hiraide:2008eu}
K.~Hiraide et~al.
\newblock {Search for Charged Current Coherent Pion Production on Carbon in a
  Few-GeV Neutrino Beam}.
\newblock {\em Phys. Rev.}, D78:112004, 2008.

\bibitem{Hayato:2002sd}
Y.~Hayato.
\newblock {NEUT}.
\newblock {\em Nucl. Phys. Proc. Suppl.}, 112:171--176, 2002.

\bibitem{Mitsuka:2008zz}
G.~Mitsuka.
\newblock {NEUT}.
\newblock {\em AIP Conf. Proc.}, 981:262--264, 2008.

\bibitem{Casper:2002sd}
D.~Casper.
\newblock {The nuance neutrino physics simulation, and the future}.
\newblock {\em Nucl. Phys. Proc. Suppl.}, 112:161--170, 2002.

\bibitem{AguilarArevalo:2007it}
A.~A. Aguilar-Arevalo et~al.
\newblock {A Search for electron neutrino appearance at the $\Delta m^{2} \sim
  1$eV$^{2}$ scale}.
\newblock {\em Phys. Rev. Lett.}, 98:231801, 2007.

\bibitem{Smith:1972xh}
R.A. Smith and E.J. Moniz.
\newblock Neutrino reactions on nuclear targets.
\newblock {\em Nuclear Physics B}, 43:605 -- 622, 1972.

\bibitem{Moniz:1971mt}
E.~J. Moniz et~al.
\newblock {Nuclear fermi momenta from quasielastic electron scattering}.
\newblock {\em Phys. Rev. Lett.}, 26:445--448, 1971.

\bibitem{AguilarArevalo:2008fb}
A.~A. Aguilar-Arevalo et~al.
\newblock Measurement of muon neutrino quasi-elastic scattering on carbon.
\newblock {\em Phys. Rev. Lett.}, 100:032301, 2008.

\bibitem{Budd:2003wb}
Howard~Scott Budd, A.~Bodek, and J.~Arrington.
\newblock {Modeling quasi-elastic form factors for electron and neutrino
  scattering}.
\newblock 2003.

\bibitem{AguilarArevalo:2010zc}
A.~A. Aguilar-Arevalo et~al.
\newblock {First Measurement of the Muon Neutrino Charged Current Quasielastic
  Double Differential Cross Section}.
\newblock {\em Phys. Rev.}, D81:092005, 2010.

\bibitem{Rein:1980wg}
D.~Rein and L.~M. Sehgal.
\newblock {Neutrino Excitation of Baryon Resonances and Single Pion
  Production}.
\newblock {\em Ann. Phys.}, 133:79, 1981.

\bibitem{Rein:1987cb}
D.~Rein.
\newblock {Angular Distribution in neutrino induced single pion production
  processes}.
\newblock {\em Z. Phys.}, C35:43--64, 1987.

\bibitem{Rein:1982pf}
D.~Rein and L.~M. Sehgal.
\newblock {Coherent $\pi^0$ Production in Neutrino Reactions}.
\newblock {\em Nucl. Phys.}, B223:29, 1983.

\bibitem{Rein:2006di}
D.~Rein and L.~M. Sehgal.
\newblock {PCAC and the Deficit of Forward Muons in $\pi^+$ Production by
  Neutrinos}.
\newblock {\em Phys. Lett.}, B657:207--209, 2007.

\bibitem{Gluck:1998xa}
M.~Gluck, E.~Reya, and A.~Vogt.
\newblock {Dynamical parton distributions revisited}.
\newblock {\em Eur. Phys. J.}, C5:461--470, 1998.

\bibitem{Bodek:2003wd}
A.~Bodek and U.~K. Yang.
\newblock {Modeling neutrino and electron scattering cross sections in the few
  GeV region with effective LO PDFs}.
\newblock {\em hep-ex/0308007}, 2003.

\bibitem{Glazov:1993ur}
A.~Glazov, I.~Kisel, E.~Konotopskaya, and G.~Ososkov.
\newblock {Filtering tracks in discrete detectors using a cellular automaton}.
\newblock {\em Nucl. Instrum. Meth.}, A329:262--268, 1993.

\bibitem{Hiraide:2009zz}
K.~Hiraide.
\newblock {\em {A Study of Charged Current Single Charged Pion Productions on
  Carbon in a Few-GeV Neutrino Beam}}.
\newblock PhD thesis, Kyoto University, 2009.
\newblock FERMILAB-THESIS-2009-02.

\bibitem{Kurimoto:2010zz}
Y.~Kurimoto.
\newblock {\em {Measurement of Neutral Current Neutral Pion Production on
  Carbon in a Few-GeV Neutrino Beam}}.
\newblock PhD thesis, Kyoto University, 2010.
\newblock FERMILAB-THESIS-2010-12.

\bibitem{Bodek:2007vi}
A.~Bodek, S.~Avvakumov, R.~Bradford, and Howard~Scott Budd.
\newblock {Extraction of the Axial Nucleon Form Factor from Neutrino
  Experiments on Deuterium}.
\newblock {\em J. Phys. Conf. Ser.}, 110:082004, 2008.

\bibitem{Rodriguez:2008eaa}
A.~Rodriguez et~al.
\newblock {Measurement of single charged pion production in the charged-current
  interactions of neutrinos in a 1.3 GeV wide band beam}.
\newblock {\em Phys. Rev.}, D78:032003, 2008.

\bibitem{Kitagaki:1986ct}
T.~Kitagaki et~al.
\newblock {Charged current exclusive pion production in neutrino deuterium
  interactions}.
\newblock {\em Phys. Rev.}, D34:2554--2565, 1986.

\bibitem{Day}
D.~Day et~al.
\newblock Study of neutrino d charged current two pion production in the
  threshold region.
\newblock {\em Phys. Rev.}, D28:2714--2720, 1983.

\bibitem{teppei:2007ru}
A.~A. Aguilar-Arevalo et~al.
\newblock {Measurement of muon neutrino quasi-elastic scattering on carbon}.
\newblock {\em Phys. Rev. Lett.}, 100:032301, 2008.

\bibitem{AguilarArevalo:2010bm}
A.~A. Aguilar-Arevalo et~al.
\newblock {Measurement of Neutrino-Induced Charged-Current Charged Pion
  Production Cross Sections on Mineral Oil at E$_{\nu}\sim 1~\textrm{GeV}$}.
\newblock {\em Phys. Rev.}, D83:052007, 2011.

\bibitem{AguilarArevalo:2011sz}
A.~A. Aguilar-Arevalo et~al.
\newblock {Measurement of the neutrino component of an anti-neutrino beam
  observed by a non-magnetized detector}.
\newblock {\em arXiv:1102.1964}, 2011.

\bibitem{Nakahata:1986zp}
M.~Nakahata et~al.
\newblock {Atmospheric Neutrino Background and Pion Nuclear Effect for Kamioka
  Nucleuon Decay Experiment}.
\newblock {\em J. Phys. Soc. Jap.}, 55:3786, 1986.

\bibitem{Sjostrand:1993yb}
T.~Sjostrand.
\newblock {High-energy physics event generation with PYTHIA 5.7 and JETSET
  7.4}.
\newblock {\em Comput. Phys. Commun.}, 82:74--90, 1994.

\bibitem{Nakamura:2010zzi}
K.~Nakamura and Particle~Data Group.
\newblock Review of particle physics.
\newblock {\em Journal of Physics G: Nuclear and Particle Physics},
  37(7A):075021, 2010.

\end{thebibliography}

\end{document}